%% file: main.tex
\documentclass[sigconf]{acmart}

\setcopyright{none}
\renewcommand\footnotetextcopyrightpermission[1]{}
\usepackage{setspace}
\usepackage{ulem}
\usepackage{xcolor}
\usepackage{hyperref}
\usepackage{multirow}
\usepackage[utf8]{inputenc}
\usepackage[ruled, lined, longend]{algorithm2e}
\usepackage{amsmath,multicol}
\usepackage{subcaption,textcomp,comment,listings,wrapfig}

\usepackage{booktabs}
\usepackage{multirow,enumitem}

\usepackage{array}
\newcolumntype{L}[1]{>{\raggedright\let\newline\\\arraybackslash\hspace{0pt}}m{#1}}
\newcolumntype{C}[1]{>{\centering\let\newline\\\arraybackslash\hspace{0pt}}m{#1}}
\newcolumntype{R}[1]{>{\raggedleft\let\newline\\\arraybackslash\hspace{0pt}}m{#1}}

\newcommand{\sys}{\texttt{Deck}\xspace}
\newcommand{\gateway}{\textit{Coordinator}\xspace}
\newcommand{\executor}{\textit{Execution Sandbox}\xspace}

\newcommand{\distance}{10pt}
\begin{document}
    \title{Device-centric Federated Analytics At Ease}
	\author{Li Zhang, Junji Qiu, Shangguang Wang, Mengwei Xu}
	\affiliation{
		\institution{Beijing University of Posts and Telecommunications}
		\country{Beijing, China}
	}
	\input{sec-abstract}
	\maketitle
	\pagestyle{plain}
	\input{sec-intro}
	\input{sec-design}
	\input{sec-impl}
	\input{sec-eval}
	\input{sec-related}
	\input{sec-discuss}
	\input{sec-conclusions}
	\bibliographystyle{plain}
	\bibliography{ref}
\end{document}

%% file: sec-abstract.tex
\begin{abstract}
    Nowadays, high-volume and privacy-sensitive data are generated by mobile devices, which are better to be preserved on devices and queried on demand.
	However, data analysts still lack a uniform way to harness such distributed on-device data.
	In this paper, we propose a data querying system -- \sys, that enables flexible device-centric federated analytics.
	The key idea of \sys is to bypass the app developers but allow the data analysts to directly submit their analytics code to run on devices, through a centralized query coordinator service.
	\sys provides a list of standard APIs to data analysts and handles most of the device-specific tasks underneath.
	\sys further incorporates two key techniques:
	(i) a hybrid permission checking mechanism and mandatory cross-device aggregation to ensure data privacy;
	(ii) a zero-knowledge statistical model that judiciously trades off query delay and query resource expenditure on devices.
	We fully implement \sys and plug it into 20 popular Android apps.
	An in-the-wild deployment on 1,642 volunteers shows that \sys significantly reduces the query delay by up to 30$\times$ compared to baselines.
	Our microbenchmarks also demonstrate that the standalone overhead of \sys is negligible.
\end{abstract}

%% file: sec-intro.tex
\section{Introduction}\label{sec:intro}

Data is perhaps the most valuable asset in our digital world nowadays.
This is especially the case for data generated by mobile and IoT devices, which help reveal user behaviour and physical worlds changes~\cite{han2007frequent,liu2012mining}.
To analyze data generated by massive devices, the mainstream methodology is \textit{cloud-centric}~\cite{han2007frequent}, where the devices stream the data to clouds so data users\footnote{In this paper, we term those who query mobile data as \textit{data users}.
A data user can be a human (e.g., data analyst) or a running program (e.g., a recommendation system that relies on realtime user preferences for online decision~\cite{DBLP:conf/www/Al-GhosseinAB18,DBLP:conf/recsys/DiasLLEL08,liu2019hydra}).
To be noted, \textit{app developers}, who build the app logic and data querying system, are often different from data users, e.g., they may come from different departments of the same company, or even from different companies for the scenario of cross-entity data sharing~\cite{cross-entity-data-sharing}.
}
can analyze them with unified frameworks like Spark~\cite{spark2012nsdi} and Flink~\cite{carbone2015flink}.

\textbf{Data staying on devices}, however, is an emerging paradigm in recent years.
The reasons are twofold.
First, with the ever-growing public concerns over data privacy and relevant regulations (e.g., GDPR~\cite{gdpr} and CCPA~\cite{ccpa}),
companies can no longer collect data arbitrarily as they used to.
Second, the data volume is explosively growing, outpacing the upgrading speed of network capacity, especially the wide-area network (estimated to be 4$\times$ and 1.5$\times$ from 2018 to 2022, respectively~\cite{statista-mobile-traffic,cisco-white-paper}).
For instance, a recent work~\cite{xu2021video} shows 99\% videos captured by a campus' IoT cameras will not be queried and building an analytics system that retrieves data on demand can reduce the network cost by two orders of magnitudes.
Such private and cold data are better to be preserved on devices.
Only a small, critical portion of them shall be permitted (by privacy or network) to be streamed to clouds.

\input{fig-interaction}

The trend catalyzes \textit{device-centric federated analytics}~\cite{google-blog-fa}, as the data processing is shifted from clouds to devices and only privacy-friendly or more compact results are returned to clouds.
Use cases abound:
(1) Google, the early proposer of federated analytics, employs it in Pixel phones to identify the songs playing in a region with an on-device database previously distributed~\cite{google-playing-now}.
It enables Google engineers to improve the song database, e.g., 5\% accuracy improvement globally as reported, without any phone revealing which songs were heard.
(2) Mobile input methods typically record what user types on the keyboard on local storage for those who opt in~\cite{apple-dp}.
As Figure~\ref{fig:interaction} shows, at any time, a sociologist may issue a query to characterize the users' input patterns, e.g., the most popular emoji used last week, to understand the society-scale reactions to certain public events.
(3) Federated learning~\cite{bonawitz2019towards} is a special paradigm of device-centric federated analytics, where the devices collaboratively train a machine learning model without sharing the raw training data.
Meanwhile, its testing process, i.e., to measure how an FL-trained model performs in the real world, is another important federated analytics scenario.
For all above cases, keeping data on local storage not only maximizes privacy preservation, but also enables fine-grained per-user privacy configuration, e.g., each user could turn off their data accessibility at any time.

However, there lacks a flexible and uniform framework to perform device-centric federated analytics.
The status quo is ad-hoc:
(1) The app developers \textit{hard-code} the analytics logic into the specific app to get post-analytics results to clouds;
(2) A new query needs to undergo a \textit{complete software engineering} process, i.e., app developers need to revise the app's logic, run testing cases, and upgrade the app on devices, even though the query is quite simple.
Such a process is labor-intensive, cumbersome, and inefficient~\cite{liu2014imashup}.
More advanced tools~\cite{noauthor_tencenttinker_2021,noauthor_andfix_2021,noauthor_robust_2021,noauthor_alibabadexposed_2021} relieve the developers from rebuilding the whole applications, but have not fundamentally addressed the issue of programmability in federated analytics.

As a novel concept, very few literature explores the field of federated analytics (FA)~\cite{fa-survey}.
They mostly build prototype frameworks to demonstrate its feasibility, e.g., video analytics on network cameras~\cite{hu2021feva}, collaborative frequent pattern mining~\cite{infocom22-fedfpm} or tackling data skewness in FA~\cite{wang2021fedacs}.
They are either designed for specific apps or solving dedicated type of problems, rather than providing a unified solution to perform FA in the wild.
TensorFlow Federated~\cite{tff} doesn't provide an end-to-end solution for FA on mobile devices.
Earlier, there are closely related concepts to federated analytics, e.g., in-sensor-network processing~\cite{yao2003query,madden2005tinydb}, wide-area-network analytics~\cite{pu2015low,viswanathan2016clarinet}, and distributed differential privacy analytics~\cite{roth2019honeycrisp,roth2020orchard,roth2021mycelium}.
Those systems do not fit into our proposed scenarios for one or many of the following reasons:
(1) They may ignore the privacy aspect of data query, which is a primary incentive to decentralize the data analytics.
(2) They mostly target traditional SQL query optimizations, yet modern data analytics requires more flexible programmability, e.g., deep learning training or inference.
(3) They either optimize for energy consumption (on sensors) or query speed (on edge servers), 

To fill the gap, we propose \sys, the first-of-its-kind uniform framework for modern device-centric federated analytics.
The key idea of \sys is to bypass the app developers, but allow data users to directly submit their code to run on devices.
\sys provides a set of standard interfaces to data users mainly in \texttt{Java}, including \texttt{Pandas}-like DataFrame, ML-related operations, databases access, as well as Android-native function wrappers.
Underneath, \sys handles most of the device-specific tasks transparently, thus allowing data users to focus on the data analytics logic as if the data has been already retrieved and locally stored.
\sys is designed for \textit{approximate data query}~\cite{chakrabarti2001approximate,acharya1999join,babcock2003dynamic} that requires only a subset of devices to obtain a fairly confident result as shown in Figure~\ref{fig:interaction}.

\textbf{Scope of this work}
As a new paradigm in data querying systems, there emerges manifold research questions to be solved.
As the very early attempt, this work only probes into the sound architectural design of device-centric FA systems, and how it can possibly enhance the data analytics flow of existing apps.
Technically, this work targets the fundamental demands in building a data querying system as traditional distributed/SQL systems~\cite{wang2021ownership}:
(1) the querying performance, e.g., the query latency and resource cost.
(2) the data privacy guarantee.

\sys consists of two core modules:
(i) a central \gateway that receives queries from data users, generates and schedules tasks to devices, retrieves and aggregates the results from the devices, and finally returns the aggregated result to data users;
(ii) an \executor that runs on devices and handles the tasks dispatched by \gateway.
To use \sys, the app developers only need to embed the \executor into the app as an independent service,
and deploys the \gateway on cloud with a user bookkeeping system to manage how different data users can utilize the data, e.g., accessible datasets, maximal query frequency, etc.

\textbf{Key techniques}
\sys's design in using a central coordinator for device dispatching and results collection also offers the opportunity to employ a consolidated policy to manage data usage.
To design such a policy, however, we face two primary challenges.

$\bullet$ First, how to offer data users flexible programmability while guaranteeing data privacy?
Since the analytics logic could be far more complicated than a SQL query as shown in Table~\ref{tab:apps}, \sys is designed to allow almost arbitrary \texttt{Java} operations.
In our threat model, \gateway and \executor are trusted, while data users can be curious and malicious, therefore various attacks can be potentially performed during a query, e.g., permission outreach through Java reflection.

($\S$\ref{sec:privacy}) \sys employs multiple mechanisms to guarantee data privacy.
To defend permission outreach, \sys combines (i) a lightweight \textit{Annotation and Proxy} mechanism powered by \sys's defined APIs that all resource accessing operations will be redirected to;
(ii) a hybrid code analysis technique to detect prohibited method usage in a query.
Through above techniques, \sys protects sensitive data and APIs from unauthorized data users at both Java and native code.
To defend differential analysis attacks, \sys enforces a query to use cross-device aggregation, so data users can access the statistical patterns of the queried dataset but not personal information.

$\bullet$ Second, how to trade off resource expenditure and query delay?
An analytic query needs to involve enough number of devices (specified by data users in \sys) to be statistically meaningful.
However, the query response time of every single device is volatile due to the dynamic network environment and device status as we experimentally demonstrate in $\S$\ref{sec:query-delay-measurements}.
Therefore, sending a query only to the required number of devices often results in an unpredictable and long query delay.
To mitigate such impacts from devices' dynamic status, it is straightforward to redundantly schedule a query to more devices than expected.
Such redundancy, however, consumes precious hardware resources (e.g., network, compute and energy), therefore needs to be minimized.

($\S$\ref{sec:scheduling})
To better trade off the resource expenditure and query delay, \sys incorporates a statistical model to guide the device dispatching strategy in an \textit{incremental manner}.
The key idea is that, based on the progress of results being returned, we can calculate the expectation of the remaining time to finish a query.
For delayed queries, \sys can opportunistically dispatch the analytics tasks to more devices to compensate for the delay from slow devices.
The policy requires zero apriori knowledge about devices' hardware status and incurs no additional privacy or network cost.

\textbf{Field deployment and evaluation}
We fully implemented \sys and embedded it into 20 popular open-source Android apps with diversified data queries.
We conducted a field deployment by installing 3 \sys-enabled apps on 1,642 real users' mobile phones for 14 days.
Through 3,517 queries and 232,779 responses from those devices, \sys shows superior performance: its 99th-percentile query delay is only 2 seconds on average, which is up to 30$\times$ smaller than using a fixed redundancy.
Besides, developing with \sys is highly efficient, as one query requires only 77 lines of code on average across the 20 apps we built.
Compared to industrial hot-fix libraries~\cite{noauthor_tencenttinker_2021}, \sys is up to 105$\times$ faster in dispatching a query for its decoupled query and app logic.

\noindent \textbf{Contributions} are summarized below.
\begin{itemize}[leftmargin=*]
    \item We propose a uniform framework, \sys, for device-centric federated analytics in a swift, flexible manner.
    \sys provides a list of easy-to-use programming interfaces to data users and hides the device-specific details.
    \item We address two key challenges to make \sys safe and efficient.
    One is through a comprehensive, hybrid security mechanism to ensure data privacy.
    The other one is through a statistical model that incrementally involves more devices for fast response.
    \item We implement \sys, plug it into popular Android apps, and deploy those apps on 1,642 users' devices to study their performance in the wild.
    The results demonstrate \sys's effectiveness in delivering fast query responses with little resource waste.
\end{itemize}

%% file: fig-interaction.tex
\begin{figure}
	\centering					
	\includegraphics[width=0.4\textwidth]{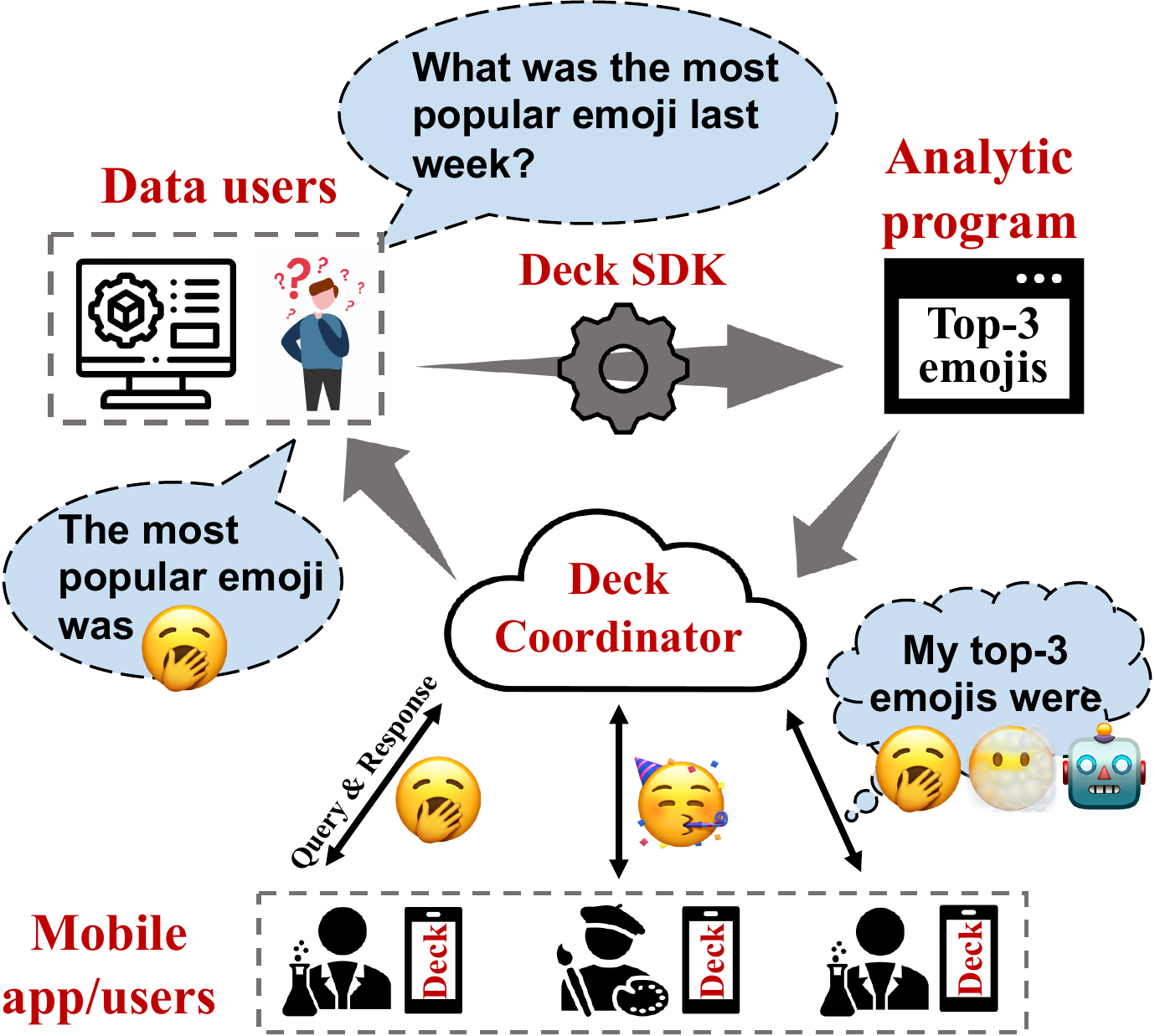}
	\caption{A use case of performing federated analytics.
	}
	\label{fig:interaction}
\end{figure}

%% file: sec-design.tex
\input{tab-APIs}

\section{Deck Design}\label{sec:design}

\input{sec-design-goals}

\input{sec-design-deploying-deck}

\input{sec-design-usecase}

\input{sec-design-workflow}

\input{sec-privacy}

\input{sec-scheduling}

%% file: tab-APIs.tex
\begin{table*}[t]
	\centering					
	\includegraphics[width=0.95\textwidth]{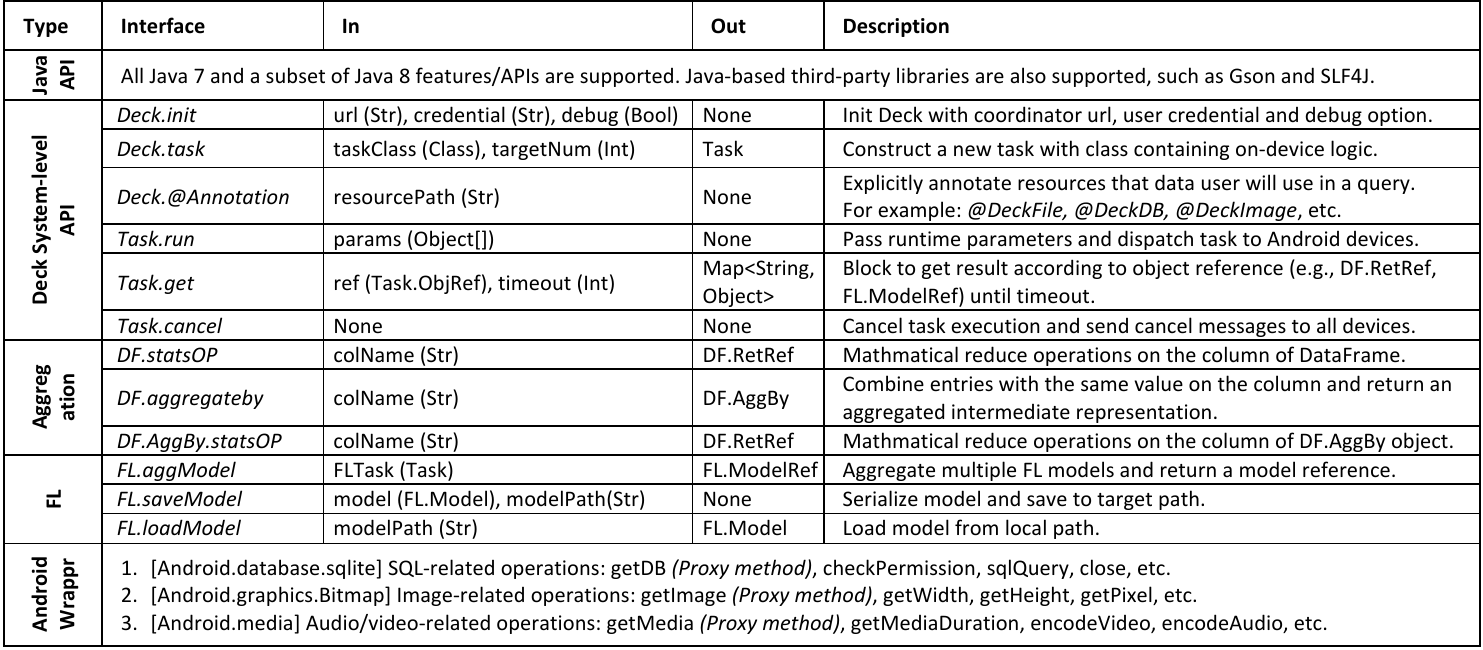}
	\caption{\sys's primary APIs provided to data users.}
	\label{tab:APIs}
	\vspace{-10pt}
\end{table*}

%% file: sec-design-goals.tex
\subsection{Design principles and goals}

When designing \sys, we make the following assumptions about data users.

\noindent $\bullet$
They have very little knowledge of mobile programming, but only standard data analytics languages, i.e., Java in our case to facilitate query execution on Android devices.
Note that Java has extensive usage in big data processing frameworks like Spark~\cite{spark2012nsdi} and Flink~\cite{carbone2015flink}.

\noindent $\bullet$
They care little about the resource cost to run the query on devices.
Somewhat paradoxically, they expect the query delay to be as short as possible, because a long delay annoys them, especially when the query is part of an online service~\cite{DBLP:conf/www/Al-GhosseinAB18,DBLP:conf/recsys/DiasLLEL08,liu2019hydra}.

\noindent $\bullet$
They have zero incentive to guarantee the privacy validity of their queries.
In reality, the data users may be careless or even malicious, as the data might be shared among different parties to maximize their business returns.

Therefore, \sys needs to hide the device-specific details, trade off the resource costs and query delay, and impose privacy-preserving mechanisms in an automated and non-intrusive manner.
It is achieved through its overall architectural design, API interfaces, and the key techniques presented in the following subsections.

Federated analytics is mostly \textit{approximate}~\cite{chakrabarti2001approximate,acharya1999join,babcock2003dynamic}, meaning it does not require a complete dataset (from all devices) to acquire the final result; and \sys is designed for such approximate queries.
Such queries are prevalent, as (1) iterating over each device can be exceptionally slow; 
(2) a small subset of data is often enough to generate a confident result~\cite{agarwal2013blinkdb,fan2014querying,fan2015querying}.
According to our query to study the average time interval between users' typing sequences (Q1 in Table~\ref{tab:apps}), we find that querying 30 users gives us a 10\% confidence interval 
with a 95\% confidence level.

%% file: sec-design-deploying-deck.tex
\subsection{Deploying Deck}

Deploying \sys consists of two core stages.

\noindent $\bullet$ First, app developers need to plug \sys's device-side library into their apps and pair it with local datasets.
We call them \sys-enabled apps.
App developers may involve as many datasets that could be potentially used for analytics as possible, and \sys employs a user-dataset permission control system to allow a user to access only certain datasets with granted permission.
A \sys-enabled app will maintain a persistent connection with the central \gateway, which is usually deployed on a cloud server, for instant task/result message exchanging.

\noindent $\bullet$  Second, app developers deploy a central \gateway on a server that handles queries from data users and interacts directly with devices for task dispatching.
It works as an intermediary between data users and the devices.
Such a centralized design opens huge space for a unified design to trade off the query delay and on-device resource costs, as well as imposing the query privacy validation.
The scalability of \sys, e.g., to more data users or devices, can be effectively expanded by scaling the hardware capacity of the servers where \textit{Coordinators} are deployed.

%% file: sec-design-usecase.tex
\subsection{APIs and use cases}\label{sec:fl-use-case}

\sys's programmability is based on Android-compatible Java features\footnote{Android supports all Java 7 and most Java 8 features.}.
It also means the data users can use any compatible third-party Java libraries.
Besides, data users can also use JNI technology for programming with native C++ code, as used in our federated learning query to facilitate on-device model training.

In addition, \sys provides a set of simple programming interfaces (shown in Table \ref{tab:APIs}) to facilitate data users to write data queries:
(1) \sys's system-level APIs that facilitate data users to dispatch and manage the analytic tasks.
An analytic task is specified with a given number of devices to run through \textit{Deck.task}.
The execution function \textit{Task.run} is non-blocking until the \textit{Task.get} is invoked.
(2) \sys's application-level APIs that run on the \gateway, e.g., \textit{mean, sum, aggregateby} for DataFrame and \textit{FL.aggModel} for FL, etc.
We retrofit the design of \texttt{Pandas}~\cite{mckinney2010data} and \texttt{MNN}~\cite{mnn} to improve the usability of these APIs.
(3) \sys's Android wrappers for processing different types of data such as databases, images, and media.
Those APIs for Android platform are typically implemented more efficiently as compared to Java APIs if there exist.
They also implement \sys's \textit{Annotation and Proxy} mechanism, responsible for permission checking and avoiding out-of-boundary resource accessing by data users, who are prohibited to use the origin Android APIs directly.
All above APIs are packed into a Java SDK for data users, with which data users only need to focus on implementing the data analytic logic.

\input{code-fl}

\textbf{A case of federated learning}
is shown in Listing \ref{code:fl}.
In the code snippet, \texttt{Class OnDeviceActor} contains the analytic logic to be run on devices, 
and \texttt{Class DeckSys} contains the management logic using \sys's system-level APIs, which will be executed on data user's desktop.
This program first disables debug mode and connects to \gateway with credential file using \textit{Deck.init}.
In each round, we construct a \textit{Task} object using \textit{Deck.task}.
The task will be dispatched to devices by \gateway once \textit{task.run} is called and all local parameters will be transferred to \gateway.
We can use \sys's FL API to invoke model aggregation across trained models returned from each device by passing a \textit{Task} object as an identifier.
Once getting an aggregated model using \textit{task.get} from the \gateway, we can test the accuracy or run other FL-related operations upon it locally before entering the next training round.
At last, we save the trained model to our local desktop.

%% file: code-fl.tex
\definecolor{dkgreen}{rgb}{0,0.6,0}
\definecolor{gray}{rgb}{0.5,0.5,0.5}
\definecolor{mauve}{rgb}{0.58,0,0.82}

\lstset{frame=tb,
  language=Java,
  aboveskip=3mm,
  belowskip=3mm,
  showstringspaces=false,
  columns=flexible,
  basicstyle={\footnotesize\ttfamily},
  numbers=none,
  numberstyle=\tiny\color{gray},
  keywordstyle=\color{blue},
  commentstyle=\color{dkgreen},
  stringstyle=\color{mauve},
  breaklines=true,
  breakatwhitespace=true,
  tabsize=3,
  literate={\ \ }{{\ }}1,
}

\begin{lstlisting}[caption={A federated learning query.},abovecaptionskip=10pt,captionpos=b,label={code:fl}]
class OnDeviceActor extends Actor {
    @DeckFile("datasetPath")
    public static Result<Model> run(Model model, int epoch) {
        Optimizer optimizer = new Optimizer("SGD");
        for (int i = 0; i < epoch; ++i) {
            model = trainIter(model, optimizer, "datasetPath");
              // Train model using MNN native method
        }
        return new Result<>(model);
    }
}
public class DeckSys {
    public static void main(String[] args) {
        Deck.init("ip:port", "credential", false);
        Model model = FL.loadModel("initModel");
        for (int i = 0; i < 40; ++i) {  // Rounds
            Task task = Deck.task(DeviceTrain, 50); 
                // Aggregate models from 50 devices
            task.run(model, 5);  // Non-blocking invoke
            Fl.ModelRef modelRef = FL.aggModel(task);  
                // Get model reference immediately
            model = task.get(modelRef, 60*10)["model"];
                // Blocking to get aggregated model
        }
        FL.saveModel(model, "trainedModel");
    }
}
\end{lstlisting}
\vspace{-10pt}

%% file: sec-design-workflow.tex
\subsection{Workflow}

\input{fig-workflow}

Figure~\ref{fig:workflow} shows how \sys works internally.

\textbf{Local compiling}
A query begins with data users running the query on their local machine, e.g., a desktop.
The device-side query code will be compiled to Java-Class files locally, then uploaded to \gateway with data user's credential, third-party dependencies, and the parameters.
All the processes will be performed automatically by \sys's Data-user SDK.

\textbf{User bookkeeping}
Upon receiving a query request, the \gateway first authenticates the data user and checks if there is still enough ``quantum'' left for her.
The quantum is defined as how many devices have been queried by this data user in a past period, e.g., a month.
The use of quantum is to prevent data users from excessively submitting queries, and potentially be used in a billing model (not the focus of this work).

\textbf{Privacy pre-checking}
\gateway then performs permission checking on the submitted query after compiling the submitted Java-Class files to \texttt{dex} files in two steps:
static permission checking to find the prohibited API usages and violated data access operations;
dynamic code injection to monitor the runtime behavior on devices.
The details are discussed in $\S$\ref{sec:privacy}.

\textbf{Task scheduling}
\sys begins to dispatch the revised query to devices for execution.
These devices are selected from a device pool that \gateway maintains for those available devices.
In order to optimize resource expenditure and query delay, a task scheduling policy is needed to determine \textit{how many} and \textit{what} devices the task should be dispatched to.
\sys now employs a statistical model as described in $\S$\ref{sec:scheduling}.
Not all queries will be dispatched to real devices -- \sys allows the query to run in \textit{debug} mode specified in the \texttt{Deck.init} API, so the query will only run on \gateway with dumb data.
It facilitates data users to test their code efficiently.

\textbf{On-device execution}
\sys's execution sandbox runs a task in a low thread priority to avoid compromising user experience in using the mobile device.
The task is executed till completion unless:
(i) the permission inspector is notified of a runtime permission violation by the code injected at \gateway;
(ii) the device is notified by the \gateway that the query is completed with enough results or canceled by the data user.

\textbf{Results aggregation}
The FA results returned from devices will be transmitted to \gateway and aggregated across devices.
Such aggregation is performed continuously with incoming results in a non-blocking manner to reduce the aggregating latency.
When the results meet the targeting number, it returns the aggregated results to data users.
Such post-aggregation data is often much less privacy-sensitive and more compact than the raw data.

%% file: fig-workflow.tex
\begin{figure*}[t]
	\centering					
	\includegraphics[width=0.95\textwidth]{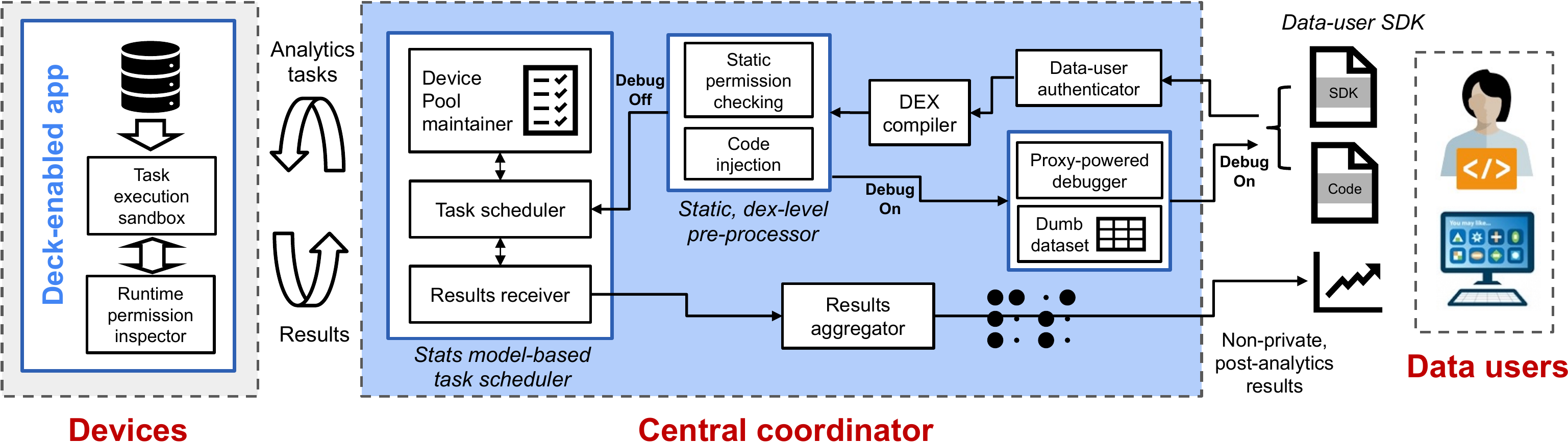}
	\caption{The workflow of \sys.}
	\label{fig:workflow}
\end{figure*}

%% file: sec-privacy.tex
\section{Privacy Guarding}\label{sec:privacy}

\input{sec-threat-model}

\subsection{Defending permission outreach}
While \sys allows data users to write almost arbitrary Java code, somehow the access is restricted for privacy preservation in two aspects: data (databases, images, etc.) and APIs (Android geolocation services, audio recording, etc.)
\sys needs to guarantee that the unqualified queries not satisfying those policies will be rejected, and deal with potential intentional attacks like bypassing static code checking through Java reflection mechanism.
The design matches Android's permission system that aims to protect access to restricted data and actions~\cite{android-permission}.

\subsubsection{Design choices}
We first discuss existing techniques that can potentially defend the permission outreach.

\noindent \textbf{In-app privilege isolation}~\cite{qu2017dydroid,seo2016flexdroid} has been recently explored to provide a separated, controllable permission environment for dynamically loaded code from the original context.
However, this approach requires modification to the Android framework, while \sys is an app-level framework to be paired with existing Android apps, thus making it undoable.

\noindent \textbf{Proxy mechanism}~\cite{sohan2015spyrest,hong2001webquilt} is widely used in software testing, especially for web applications~\cite{spring-mock-test,go-testify}. 
The key idea of proxy mechanism is to redirect privacy-sensitive methods to self-defined privacy-guaranteed methods.
We can leverage \sys's programming APIs to implement this conveniently. 

\noindent \textbf{Static code analysis}~\cite{livshits2005finding} technique can detect the direct usage of Java classes and methods, but the malicious attackers can still bypass it through Java reflection.
With the most advanced approach, it is still challenging to restore the concrete inputs to a given function~\cite{landman2017challenges}, i.e., the class/method name in our case.
On the other hand, simply disabling the use of Java reflection leads to the unavailability of many third-party libraries (e.g., \textit{kryo} reflects the "sum.misc.unsafe" class to speed up serialization~\cite{java-unsafe-reflection}), which can severely compromise the programming flexibility of \sys.

\noindent \textbf{Dynamic code analysis} is a way to intercept concerned APIs and obtain the low-level runtime information, e.g., the classes/methods being reflected~\cite{ball1999concept,artho2004jnuke,andreasen2017survey,sun2018efficient}.
A plausible solution is to deploy the intercepted systems on our own devices or emulators along with \gateway for dynamic analysis.
This approach, however, blocks the task scheduling because the tasks must be dispatched after the checking is completed, therefore introducing additional query delay.
More than that, the code coverage of dynamic analysis is still limited and the malicious code might be bypassed~\cite{or2019dynamic}.

\subsubsection{Protecting data access}\hfill\\
\indent \textbf{Java code}
\sys employs a lightweight \textit{Annotation-Proxy} mechanism~\cite{spring-mock-test,go-testify} powered by the design of \sys's APIs (Table~\ref{tab:APIs}) to avoid malicious data accessing.
More specifically, data users are required to explicitly annotate what data they will use (e.g., a dataset file or a folder of images) in their query, i.e., \texttt{@DeckFile} in Listing~\ref{code:fl}.
These annotations will be extracted and checked by \gateway.
The queries will be rejected immediately if they request to access the data that the corresponding data users do not have permission to.
Data users are also required to use \sys's APIs to access local data, which is guaranteed by \sys's technique described in the next subsection.
These methods, acting as a \textit{Proxy}, will check if the query is accessing the not-annotated data at query runtime on devices.
If so, the query will also be rejected.

\textbf{Native code}
\sys also allows data users to write or use third-party libraries with native methods, which are compiled into shared libraries.
To avoid privacy leakage through native code, \sys leverages \textit{isolatedProcess}~\cite{android-service-isolatedprocess}, a unique Android feature that enables permission isolation of certain processes from the app.
\sys puts the native code in an isolated process with zero permission, including network, read or write external storage, etc.
Instead, the native code will be redirected to \sys's proxy methods as discussed above through inter-process communication to access their granted data.
Similarly, the proxy will check if data access is permitted.

\subsubsection{Protecting API access}\hfill\\
\indent \textbf{Java code} A query may access protected APIs through direct invoking or reflection.
\sys detects direct invoking of the protected APIs through static code analysis~\cite{livshits2005finding} at \gateway.
For reflection, \sys uses dynamic code analysis~\cite{ball1999concept,artho2004jnuke} to conduct dynamic permission inspection by ahead-of-time code injection.
More specifically, \sys injects a line of code for \texttt{class} checking for every reflection usage, as shown in Listing~\ref{code:injection}.
The injected code invokes an external function of \sys when running on devices, which looks up the \texttt{class} name in the blacklist.
If matched, the device will abort the query immediately and send a violation code to \gateway.
Both the static code analysis and code injection run at \texttt{dex} level.
They require no modification to Android framework or Linux kernel, making \sys compatible with any apps and third-party libraries.
\sys disables dynamic library loading within the query, because it can bypass the code checking and injection of \sys.

\input{code-injection}

\textbf{Native code}
In native methods, the sensitive operations may access any Java method through \texttt{JNIEnv} variable, or access the system-level method to open the file or access memory directly as we described above.
We can redirect all sensitive operations to corresponding Java methods for permission checking.
However, this requires extensive work for implementing the proxy methods for all sensitive operations in \sys's API.
As a result, we only consider the feasibility and implement some of the proxy methods.
\sys also disables the method for loading a dynamic library using \texttt{dlopen} in native code.

\subsection{Defending differential analysis}
While the results returned to data users are post-analytics, they may still contain private information at a fine granularity not acceptable to be exposed to non-trusted parties.
Such possibility of information leakage can be amplified in differential cryptanalysis attacks~\cite{lai1991markov}, where an attacker repeatedly issues a query and jointly analyzes the difference of returned results with external knowledge.
For example, if an attacker can access the logs of specific websites with device identifiers, he can easily match the existing information with the query result to locate a particular device, even though the query result does not contain device-specific information.

To defend against such attacks, \sys employs a simple yet effective mechanism: \textit{mandatory cross-device aggregation}.
A valid query must end with the specified aggregation APIs, such as \textit{sum}, \textit{count}, or other aggregation operations.
Otherwise, the query will be rejected.
Therefore, the results returned to data users only reveal the statistical patterns of the queried dataset, while exposing no private information of individuals.
In federated learning, an aggregated model is commonly treated as non-private data~\cite{bonawitz2017practical} as opposed to the models trained on each device, which can be peeked into by an attacker through techniques like membership inference attack~\cite{shokri2017membership}.
\sys also requires a query to target a pre-defined minimum number of devices (10 by default).

%% file: sec-threat-model.tex
\subsection{Threat model}\label{sec:threat-model}

In our threat model, the deployed \sys runtime (including both \gateway and \executor) is trusted, while the data users may be curious or malicious.
Such a threat model is commonly employed in practice, because the developers of popular apps often represent big companies, and are legally obligated to preserve data privacy~\cite{gdpr,ccpa}.
Instead, the data users can be any third-party entities or individuals who access the data service through business cooperation or even payment~\cite{cross-entity-data-sharing}.
The design of \sys can be extended to untrusted app developers as well, i.e., by delegating the \gateway service to trustworthy governments.

It's to be noted that the architectural design of \sys inherently includes certain privacy vantages in the FA scenario:
(1) Data staying on devices allows device-owners to hold control of their accessibility, e.g., they can turn off the accessibility to their historical data;
(2) Data users cannot specify which devices to execute their analytics, thus can not map the returned results to certain end users without external knowledge.
(3) The \gateway keeps track of each query code, thus the privacy violation can be verified afterward.
Still, there is plenty of room for data users to obtain out-of-bound private information.
In this work, we consider two main categories of attacks: permission outreach and differential analysis.

%% file: code-injection.tex
\definecolor{dkgreen}{rgb}{0,0.6,0}
\definecolor{gray}{rgb}{0.5,0.5,0.5}
\definecolor{mauve}{rgb}{0.58,0,0.82}

\lstset{frame=tb,
  language=Java,
  aboveskip=3mm,
  belowskip=3mm,
  showstringspaces=false,
  columns=flexible,
  basicstyle={\footnotesize\ttfamily},
  numbers=none,
  numberstyle=\tiny\color{gray},
  keywordstyle=\color{blue},
  commentstyle=\color{dkgreen},
  stringstyle=\color{mauve},
  breaklines=true,
  breakatwhitespace=true,
  tabsize=3,
  literate={\ \ }{{\ }}1,
}

\begin{lstlisting}[caption={A code injection example to be executed on Android devices, aims to detect disabled Class usage through reflection.},captionpos=b,label={code:injection}]
String bad_class_name = "android.os.Environment";
Deck.runtime_checker.check(bad_class_name);
	// <- Injected. The class will be checked to a blacklist.
File privateDir = (File) Class
	.forName(bad_class_name)
	.getMethod("getDownloadCacheDirectory")
	.invoke(null);
\end{lstlisting}

%% file: sec-scheduling.tex
\section{Task Scheduling}\label{sec:scheduling}

\sys needs to minimize the query delay and resource expenditure -- two critical metrics to any data querying system.
In this section, we first show how this task is challenging, and then present \sys's solution.

\noindent $\bullet$ \textbf{Query delay} is the end-to-end latency between data users submitting a FA query and receiving final results.
It mainly includes three parts: network delay between data users and \gateway, the pre-processing time at \gateway, and the task scheduling time to dispatch tasks and waiting for enough results to be returned.
However, the task scheduling time often dominates the overall query delay (shown in $\S$\ref{sec:eval}).
Therefore, we mainly focus on optimizing this part of delay.

\noindent $\bullet$  \textbf{Resource redundancy} refers to how much on-device resources have been wasted during a query.
Intuitively, the resource expenditure to run a query is multifold: network traffic, CPU cycles, battery consumption, etc.
For simplicity, we don't differentiate them but treat each device that ever runs the analytics task equally.
Therefore, we define the resource redundancy as the extra percentage of devices that run the analytics task (either finished or not) to the targeted device number: $(actual\_dev\_num/target\_dev\_num)-1$.

\input{sec-query-analysis}

\input{sec-stats-model}

%% file: sec-query-analysis.tex
\subsection{Query delay: a quick look}

\subsubsection{Preliminary Measurements}\label{sec:query-delay-measurements}

\input{fig-query-analysis}

To understand the query delay, we first conduct a preliminary measurement on the query delay based on the data described in $\S$\ref{sec:eval-settings}.

\textbf{Breakdown} The response time of each device is comprised of three parts: (1) the network delay, including task downloading and results uploading (\textit{network time}); (2) the execution time of analytics tasks on devices (\textit{exec time}); (3) the waiting time of the analytics tasks to be executed on devices (\textit{blocking time}).
Figure ~\ref{fig:query-analysis-breakdown} shows that each part contributes a nontrivial portion to the end-to-end response time.

\textbf{The long tail of response time}
A key observation is that the response time highly skews -- the 99th-MAX response time is 37,167ms and 21.5$\times$ as high as its average.
Such a long tail delay challenges the reliability of query performance.
We find the reasons are multifold:
(1) Devices with different data distribution and hardware capacity cause diverse on-device execution time.
In our federated learning query, this can cause an upmost gap of larger than 100$\times$ among devices.
(2) There are high dynamics of network delay across time.
As shown in Figure \ref{fig:query-analysis-temporal}, the average response time varies a lot within 24 hours of experiments, e.g., from 441ms to 2,397ms.
Such dynamics are prevalent in nowadays Internet, due to the wireless signal change, device movement/handoff and wide-area-network congestion, etc~\cite{mobiledynamic_mobihci11}.
(3) The device usage patterns cause the analytics tasks to be scheduled in a volatile way.
Since the device-side tasks of \sys run in low priority to not interfere with user-interactive tasks, how much CPU time it can preempt depends on OS's scheduling policies~\cite{noauthor_workmanager_nodate}.

\subsubsection{Design choices}

Scheduling analytics tasks to devices to trade off query delay and resource expenditure is a brand-new research problem.
Therefore, we first discuss several plausibly doable solutions yet are unemployed in \sys.

\textbf{One-time dispatching with fixed redundancy}
An intuitive scheduling approach is to dispatch the task to more than needed to mitigate the impacts from long-tailing devices.
For instance, Google's federated learning system uses 30\% fixed redundancy~\cite{bonawitz2019towards}.
However, one-time dispatching with a fixed redundancy is far from optimal due to the high skewness of response time across devices and time as aforementioned: 
(1) With 20\% fixed redundancy, the 99th-MAX query delay is 12,890ms, which is 14$\times$ higher than the median (920ms);
(2) With 10\% fixed redundancy, the query can't even complete within our query timeout (100 seconds).

\textbf{Discriminative task dispatching}
One may replace the random task dispatching with a discriminative one by predicting which devices are likely to respond fast.
For example, the device hardware capacity, runtime status, and network conditions can be profiled periodically and used as the indicator of device response time.
We omit such discriminative device dispatching policy for three reasons.
First, collecting device runtime information to predict the response time incurs additional privacy and network costs, more or less, which is against the original intention of this work.
Second, the response time of devices is volatile and depends on a variety of factors as aforementioned.
Those factors are often out of \sys's control and difficult to predict.
Third, dispatching tasks to devices in a discriminative may introduce statistical bias, resulting in unfaithful query results that are way off the ground truth.

%% file: fig-query-analysis.tex
\begin{figure}[t]
	\centering
	\begin{minipage}[b]{0.223\textwidth}
		\includegraphics[width=1.0\textwidth]{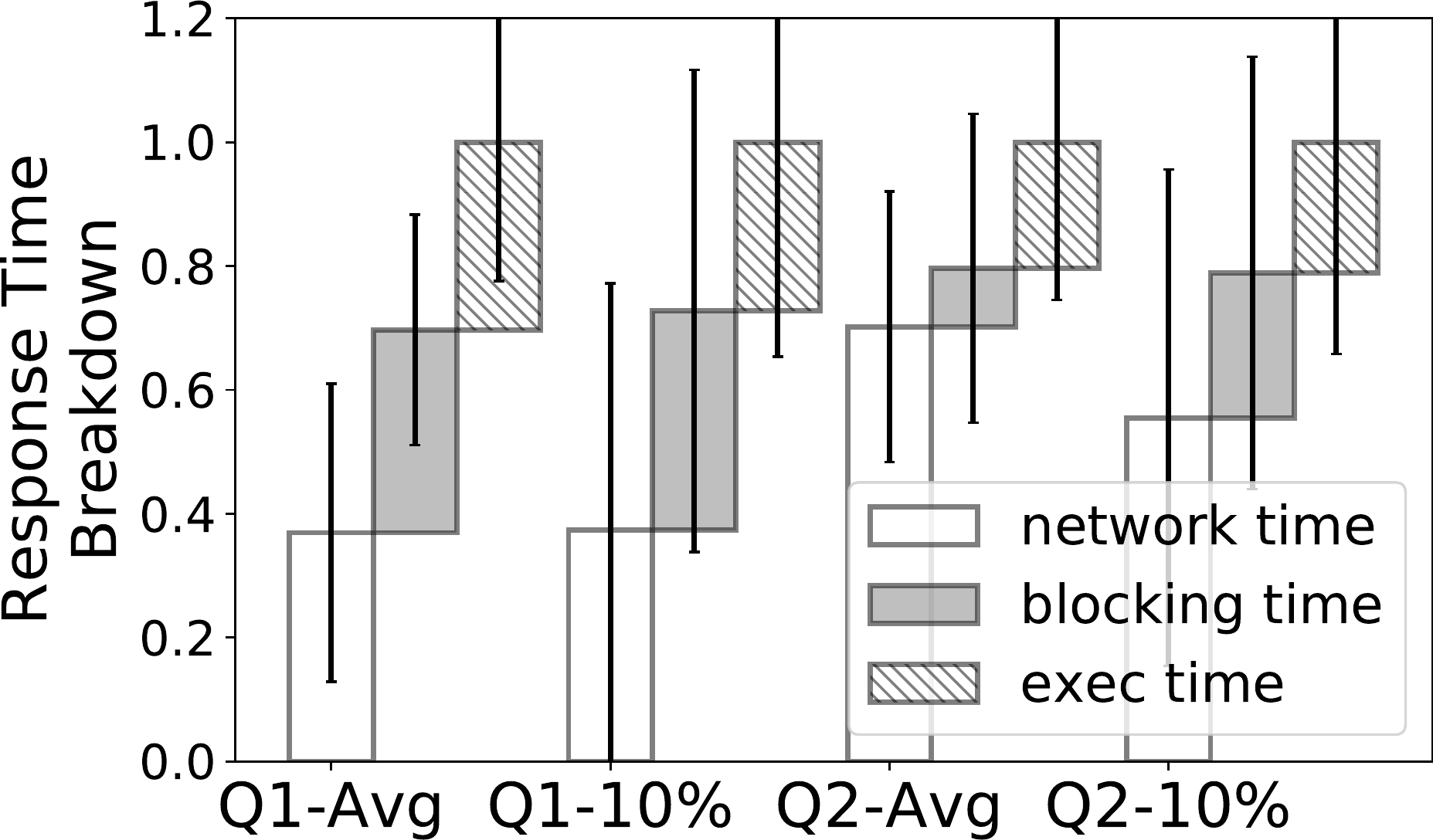}
		\subcaption{Breakdown}
		\label{fig:query-analysis-breakdown}
	\end{minipage}
	~
	\begin{minipage}[b]{0.222\textwidth}
		\includegraphics[width=1.0\textwidth]{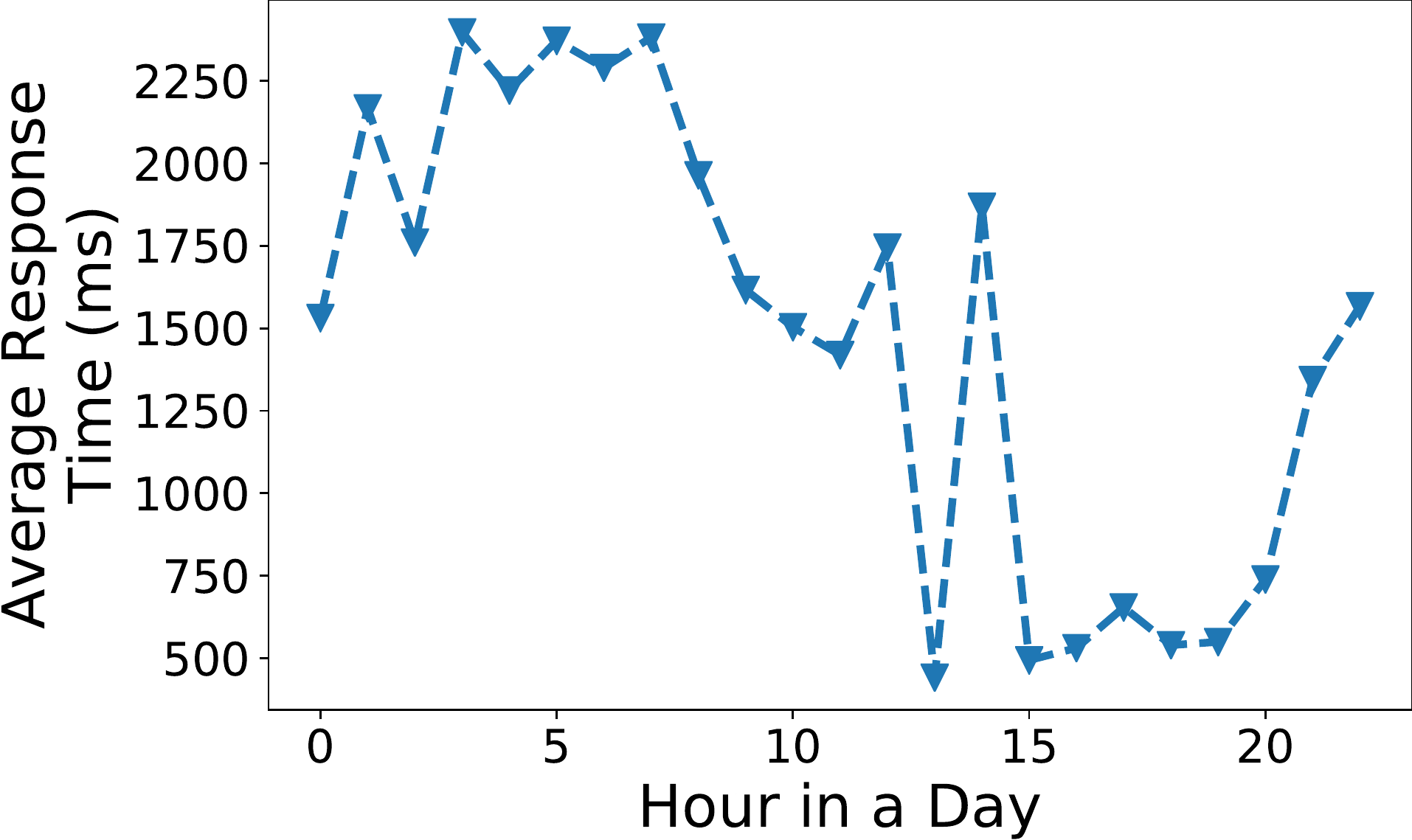}
		\subcaption{Temporal variation}
		\label{fig:query-analysis-temporal}
	\end{minipage}
	\caption{Response time analysis.}
	\label{fig:query-analysis}
\end{figure}

%% file: sec-stats-model.tex
\subsection{Zero-knowledge statistical model}\label{sec:model}

To tame the long tail of query delay, \sys is based on the idea that the query progress can be monitored on the fly, i.e., the percentage of returned result number as the query goes.
According to this, we build a statistical model, which makes \sys opportunistically dispatch the task to more devices once it finds the progress delayed.

\subsubsection{Skeleton of the model}
At the beginning of a query, \sys dispatches the analytics task to the exact number of devices as needed without redundancy.
As the query goes on and results keep coming, \sys periodically wakes up and dispatches the task to more devices (we call this wakeup interval a dispatching round).
Based on the up-to-date feedback (the number of returned results) and the response time distribution (the probability density function), we build a statistical model to determine the number of additional devices to dispatch per round.
The key idea of the statistical model is to estimate \textit{the reduction of query delay if dispatching the task to \textit{\textbf{k}} new devices}.
\sys chooses a proper \textit{\textbf{k}} to ensure that it's worth spending extra resource expenditure to speed up the query.
There are several notable design choices made by \sys.
\begin{itemize}[leftmargin=*]
	\item \sys \textit{incrementally} dispatches the analytics task to more devices instead of one-shot dispatching.
	This design can take advantage of the runtime feedbacks as the query results are continuously coming at various timestamps.
	\item \sys is \textit{knowledge-free}, as it requires zero runtime information of devices, e.g., network conditions.
	\item \sys \textit{randomly} selects devices to be dispatched, therefore no bias is introduced in the device selection stage.
\end{itemize}

\input{tab-symbols}

\subsubsection{Formulation} We propose our pseudo code in Algorithm \ref{algorithm:statistics-model} and used symbols in Table \ref{tab:symbols}.

\input{code-stats}

\textbf{Assumptions and preliminaries}
Here we assume the response time of different devices is an \textit{independent random variable} that follows certain distribution $\mathcal{N}$.
Note that we do not presume $\mathcal{N}$ to be any standard distribution like Gaussian distribution, but built it from historical queries.
Without any extra information, the expectation of a task to be returned after time slot $t$ since dispatched can be estimated by the cumulative density function (CDF) of $\mathcal{N}$, termed as $F(t)$~\cite{hogg2005introduction}.
However, at timestamp $t'$ after $t$, the \gateway is actually aware of whether the result has been returned during the time interval $[0, t']$.
If not yet, the expectation of the result being returned at time interval $[t', t]$ can be calibrated as $\frac{F(t)-F(t')}{1-F(t')}$.

\textbf{Modeling}
At the current dispatching round at timestamp $t$, \gateway has got $R(t)$ results.
If $R(t)$ is larger than the target device number $ Z$, the query completes.
Otherwise, \sys estimates how many devices it needs to send the task to.
To make this decision, we first model the expected number of the returned results $E(t_{fut})$ in a future timestamp $t_{fut}$.
Assuming that (1) we have dispatched the task to $m$ devices, among which $r = m-R(t)$ devices have not yet returned the results;
(2) the remaining $r$ devices are dispatched at timestamp $\{t_1,t_2,...,t_r\}$ ($0\le t_i < t$).
So $E(t_{fut})$ includes three parts: (1) the $R(t)$ results have been returned till time $t$; (2) the results might be returned from the remaining $r$ devices and (3) the results might be returned from the newly dispatched $k$ devices at time $t_{fut}$.
$E(t_{fut})$ can be formulated as follows:

\vspace{-10pt}
\begin{equation}
\begin{split}
E(t_{fut})&=R(t)+\sum_{i=1}^{r}{\frac{F(t_{fut}-t_i)-F(t-t_i)}{1-F(t-t_i)}}\\
&+\sum_{i=1}^{k}{F(t_{fut}-t)}
\end{split}
\label{eq:ret}
\end{equation}

\textbf{Solving}
Apparently, $E(t_{fut})$ is a monotonic function with $t_{fut}$, and we can simply use a binary search to find the $t_{fut}$ so that $E(t_{fut}) \approx Z$.
As such, $t_{fut}$ can be regarded as the expectation of the query delay.
According to Eq~\ref{eq:ret}, $t_{fut}$ relates to the number of newly dispatched device number $k$.
It's straightforward to prove that with larger $k$, the expected query delay $t_{fut}$ can be reduced.
When we dispatch the task to zero or $k$ devices, we can estimate $t_0$ and $t_k$ as task finished time 
\begin{equation} 
	E(t_0) \approx Z ;
	E(t_k) \approx Z
\end{equation}

Here, we need choose the largest $k$ so that
\begin{equation}
	\frac{t_0 - t_k}{k} \geq \eta
\end{equation}

By tuning $\eta$, we can trade off query delay and resource expenditure: lower $\eta$ favors lower delay yet higher expenditure.
For example, we record the expectation of query delay reduction and the number of devices to be dispatched in a real dispatching round in Figure~\ref{fig:delay-reduction}.
We use $\theta$ to represent how aggressive the diminished query acceleration can be when dispatching the task to one more device.
When we dispatch the task to more devices, we can see the reduction of query delay caused by each device is getting smaller.
So to avoid incurring high resource expenditure, it is not worth dispatching tasks to as many devices as possible.
For each query, we can set a manually tuned parameter $\eta$ to adjusted the trade-off between resource cost and query delay.
In each dispatching round, we can dispatch the analytics task to optimal $k$ devices where $\theta \geq \eta$, e.g., 17 in Figure~\ref{fig:delay-reduction}.
With this fine-grained constraint on dispatching device numbers, we can achieve the trade-off between query delay and resource expenditure.

\input{fig-delay-reduc-dispk}

%% file: tab-symbols.tex
\begin{table}[t]
\scriptsize
\renewcommand\arraystretch{1.3}
\begin{tabular}{lL{7cm}}
\textbf{Symbol} & \textbf{Description} \\\hline
$Z$ & Target device number specified by data users. \\\hline
$\mathcal{N}$ & Distribution of the historical response time of the query. \\\hline
$F$ & Cumulative density function (CDF) of distribution $\mathcal{N}$. \\\hline
$\eta$ & Solely tunable parameter to control the trade-off between query delay and resource expenditure. \\\hline
$t$ & Timestamp during task scheduling process. \\\hline
$k$ & Number of devices to be dispatched in a round. \\\hline
$R(t)$ & Returned results number at timestamp $t$.\\\hline
$E(t)$ & Expectation of returned results number at timestamp $t$. \\\hline
\end{tabular}
\caption{Symbols used in $\S$\ref{sec:model}.}
\vspace{-10pt}
\label{tab:symbols}
\end{table}

%% file: code-stats.tex
\begin{algorithm}
    \textbf{Initialization:}
    \begin{itemize}[leftmargin=15pt]
        \item Initial timestamp: $t=0$
        \item Device number to be dispatched: $k=0$
        \item Current results number at time $t$: $R(t)$
        \item Expectation of results number at time $t$: $E(t)$
    \end{itemize}
    \textbf{Input:}
    \begin{itemize}[leftmargin=15pt]
        \item \sys's wakeup interval: $intvl$
        \item Tuned threshold: $\eta$
        \item Target device number: $Z$
    \end{itemize}
    \textbf{Output:} Scheduling decisions during the task.\\
    \vspace{5pt}

    \setstretch{1.2}
    \While{$R(t) < Z$}{
        $t_{0}$ = $BinarySearch$($E(t_{0}) \approx Z$)\\
        \ForEach{$k_i$ in $\{k_1,k_2,...,k_n\}$}{
            $t_{k_i}$ = $BinarySearch$($E(t_{k_i}) \approx Z$)\\
            $k=max(k, k_i)$ if $\frac{t_{0}-t_{k_i}}{k_{i}} > \eta$
        }
        \If{$k>0$}{
            Dispatch the analytic task to $k$ devices
        }
        $t=t+intvl$
    }
    \caption{\textbf{\sys's task scheduling.}}
    \label{algorithm:statistics-model}
\end{algorithm}

%% file: fig-delay-reduc-dispk.tex
\begin{figure}
	\centering
	\includegraphics[width=0.27\textwidth]{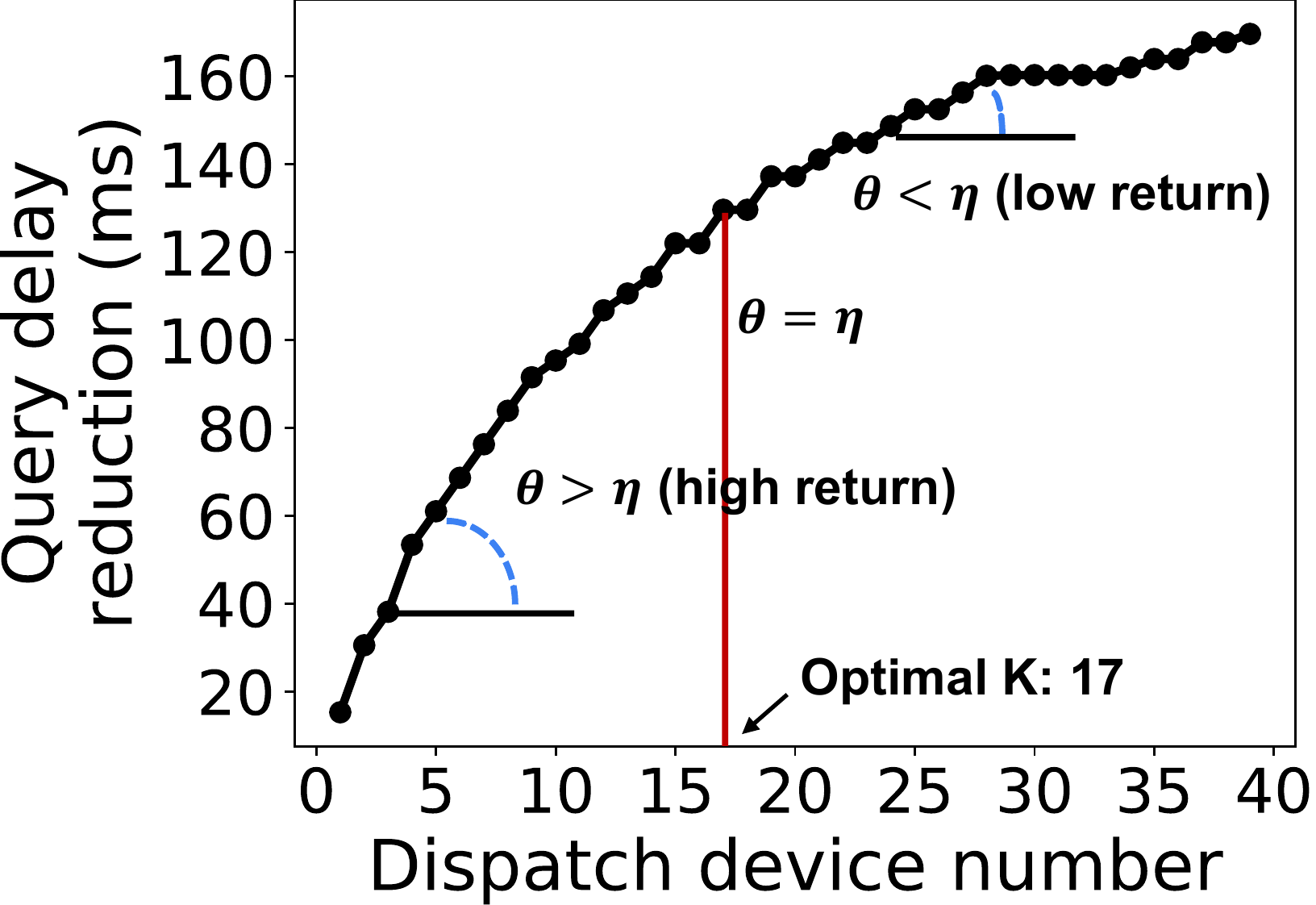}
	\caption{Diminished query acceleration with more dispatched devices.}
	\label{fig:delay-reduction}
\end{figure}

%% file: sec-impl.tex
\section{Implementation}\label{sec:impl}
We have implemented a full prototype of \sys, which mainly comprises three modules.

$\bullet$ \textbf{Coordinator} is a standalone service deployed on a cloud server and uses Redis to save all metadata at runtime for the device connection pool, task scheduling, result retrieval, cached state, etc.
Data users submit their query code to \gateway in a compiled Java-Class file format with jar dependencies, which will be further compiled into separated dex files using \textit{d8} command-line tool~\cite{d8} for caching mechanism.
The static code checking and code injection are implemented atop Soot~\cite{noauthor_soot_2021}.
In order to provide data users a uniform data format for future analysis, 
\gateway constructs query results from devices to a \texttt{Pandas}-like DataFrame using joinery~\cite{cardillo_joinery_2021}.
Later operations specified by data user's code will be performed on this DataFrame.

$\bullet$ \textbf{Android Runtime} mainly implements the device-side components of \sys, including the execution sandbox and runtime permission inspector.
They are packed into an Android library for easy integration into existing Android projects.
The communication between device and \gateway is based on WebSocket protocol using OkHttp~\cite{okhttp}.
Persistent WebSocket connections are established between \gateway and devices so \sys can maintain a device pool for task dispatching.
When receiving a task, the device will deserialize it from the message and construct a new object to represent the task.
We use Android WorkManager~\cite{noauthor_workmanager_nodate} to schedule new tasks to comply with OS's device-wise worker scheduling policy.

$\bullet$ \textbf{Data-user SDK}
provides a list of Java interfaces for data users, containing interaction APIs with \gateway, DataFrame-related operations, FL-related operations and a set of Android wrapper methods listed in Table~\ref{tab:APIs}.
It provides data users easy-to-use, concise APIs, assuming they have limited knowledge of Android programming. 

\textbf{Optimizations}
We observed that the same or similar queries are often repeatedly issued, e.g., each round of training in federated learning.
It motivates a caching mechanism to reduce the query cost.
First, the \gateway only runs the permission checker on the same dex file for one time.
Second, each device allocates a fixed storage space (20MB by default) to cache the downloaded resources.
It employs a least recently used policy for caching replacement.
Only the dex not presented locally will be retrieved from \gateway.
We evaluate the effectiveness of caching mechanism in $\S$\ref{sec:eval}.

%% file: sec-eval.tex
\input{tab-apps}

\section{Evaluation}\label{sec:eval}

\input{sec-eval-exp-settings}

\input{sec-eval-e2e-perf}

\input{sec-eval-overhead}

\subsection{Usability}\label{sec:eval-usbability}

We further study the usability of \sys as compared to a popular hot-fix library \texttt{Tinker}~\cite{noauthor_tencenttinker_2021}, which supports dex, library and resources update without reinstalling apk.
To use \texttt{Tinker}, data users need to modify the code of app and generate the patch file to be dispatched to devices.

Table~\ref{tab:tinker-overhead} summarizes the results.
For queries using third-party libraries (e.g., image processing in our example), \sys generates a larger patch file, yet is much faster than \texttt{Tinker} to compile.
This is because \texttt{Tinker} needs to compile the whole APK before generating the patch file, while \sys only needs to compile the query code.
Overall, the end-to-end response time of \sys is much shorter than \texttt{Tinker}, i.e., 40$\times$ for SQL query and 105$\times$ for image processing query.

Additionally, Table~\ref{tab:apps} also summarizes the programming efforts paid by data users when using \sys.
Overall, it only takes tens of lines of code to implement a query.

%% file: tab-apps.tex
\begin{table*}[t]
	\centering					
	\includegraphics[width=1\textwidth]{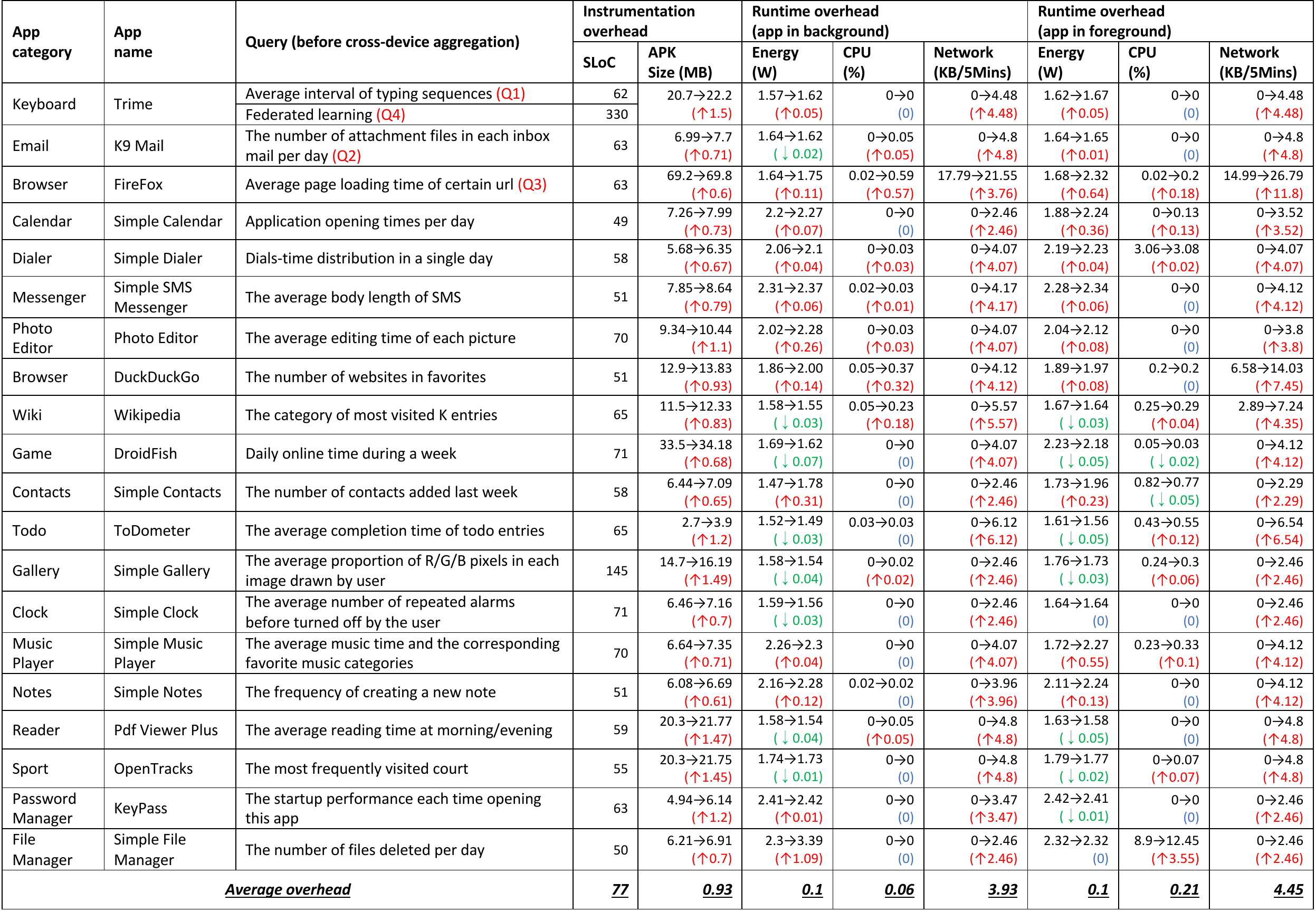}
	\caption{The apps instrumented with \sys support and their overhead without query running under controlled experiments. Here we select 20 popular open-source apps
		~\cite{app_trime,app_k9,app_fenix,app_simplecalendar,app_simpledialer,app_simplesmsmessenger,app_photoeditor,app_duckduckgo,app_wikipedia,app_droidfish,app_simplecontacts,app_todometer,app_simplegallery,app_simpleclock,app_simplemusicplayer,app_simplenotes,app_pdfviewer,app_opentracks,app_keypass,app_simplefilemanager}
		to demonstrate \sys's good practicability.
	}
	\label{tab:apps}
\end{table*}

%% file: sec-eval-exp-settings.tex
\subsection{Experiment settings}\label{sec:eval-settings}

\noindent \textbf{Apps and queries}
As shown in Table~\ref{tab:apps}, we plug \sys into 20 popular Android apps in diverse categories and design corresponding queries for each app.

\noindent \textbf{Federated learning query}
We implement a federated learning query atop \texttt{MNN}~\cite{mnn}, a lightweight DL library that supports on-device learning.
We use the standard MNIST dataset and LeNet DNN as the learning task.
According to our measurement, the training throughput of the devices involved in our field study varies from 110 to 1,040 frames/second.
The dataset is split into Non-IID and downloaded when the input method is installed for one shot.
The \texttt{MNN} library is dispatched to the user as part of \sys's query, yet it's cached during the training time according to \sys's caching mechanism in $\S$\ref{sec:impl}.

\noindent \textbf{Field deployment}
To investigate the usability and performance of \sys in the real world, especially the task scheduler, we recruited 1,642 volunteers to install three above \sys-enabled apps on their devices: input method, email client, and browser.
The volunteers were recruited through a commercial third-party crowdsourcing platform and compensated with a gift card.
To ensure the responsiveness of queries, all volunteers were asked to use these apps by default.
The \gateway is deployed on a public server with 32 CPU cores, 64 GBs memory, and 200 MB network bandwidth capacity.

The field deployment lasted for two weeks in Jul. 2021.
During the experiment, we used a long-running desktop program (as a data user) to periodically issue queries to devices for each app at an interval of 20 minutes.
The first week was treated as the data collection stage, where we issue the query to all connected devices exhaustively.
The second week was used for evaluating \sys's task scheduling policy ($\S$\ref{sec:scheduling}).
In total, we issued 3,517 queries and received 232,779 responses from devices.
We observed only a small portion of devices can respond simultaneously, as the OS often goes to sleep when device is not in use.
Among the 3 apps, the input method was the most responsive.

\noindent \textbf{Parameter settings}
By default, we set the target device number to be queried as 100, which is often enough to have a statistically meaningful result, e.g., with a bounded error less than 5\% at 95\% confidence level for Q1.
We set the wakeup interval in \sys's scheduling algorithm as 100ms for SQL query and 1,000ms for FL query.
We mostly focus on 10\% and 20\% of resource redundancy, slightly smaller than the configuration used in Google's federated learning app~\cite{bonawitz2019towards}.
We use a tighter resource budget because, in our vision, queries will be prevalent and frequently issued by many apps.
Therefore, how we can push the limits in reducing resource redundancy is of primary importance.

\noindent \textbf{Metric} For query delay, we mostly report the value of 99th-percentile (99th-MAX) delay of all queries, as it's a key metric widely adopted in data query systems~\cite{chen2021achieving}.

\noindent \textbf{Ethic considerations}
(1) This work is IRB-approved by the institution the authors are affiliated with.
(2) We only collect post-analytics data from participants that are encrypted over TLS during transmission.
(3) All data are anonymized during collection, storage, and processing.

%% file: sec-eval-e2e-perf.tex
\input{fig-eval-e2e-query-latency}

\subsection{End-to-end query delay}

\subsubsection{Breakdown}
We first break down the end-to-end query delay of two queries (Q1 and Q2) using the fixed-redundancy dispatching algorithm.
The results are summarized in Table \ref{tab:e2e-breakdown}.
Cold/warm refers to whether the caching mechanism is invoked to avoid repeated pre-processing and downloading of third-party libraries.
Overall, we find the task scheduling time dominates the query delay, contributing to more than 90\% for both cold and warm queries.
It motivates us to focus on reducing this time through a statistical model.
In addition, our caching mechanism can reduce a large amount of pre-processing time on the \gateway, for example, 322ms for Q1 and 386ms for Q2.

\input{tab-e2e-breakdown}

\subsubsection{Task scheduling performance}
We now evaluate our task scheduling policy presented in $\S$\ref{sec:scheduling}.
We compare it with the two following baselines.
\noindent $\bullet$ \textbf{OnceDispatch} simply dispatches task to devices with a fixed redundancy at the beginning of the query, which is used by traditional federated learning system~\cite{gboard-fl}.

\noindent $\bullet$  \textbf{IncreDispatch} borrows the idea from \sys to incrementally dispatch tasks according to the response from devices.
The difference is that this baseline is not guided by a statistical model but the task scheduler will wake up and check how many results does it need to complete the analytics task each interval.
Then a corresponding number of tasks will be incrementally dispatched to devices.
We empirically choose the best parameters for this baseline.

Figure \ref{fig:eval-e2e-query-latency} illustrates the 99th-MAX query delay with two fixed redundancy of 10\% and 20\%. 
A key observation is that \sys significantly reduces the query delay as compared to the baselines.
In Q1 and Q2, tasks can't complete with 10\% redundancy using \texttt{OnceDispatch} because the ultra-slow devices are dropped out in our experiments by a 100s timeout.
Its degraded performance is due to the volatile, long-tail response time as observed in $\S$\ref{sec:scheduling}.
In the most optimal case, \sys reduces the query delay by 31.65$\times$ (from 63.3s to 2s) in Q3.
\sys also outperforms \texttt{IncreDispatch} by 1.12$\times$--1.45$\times$ reduced delay for different queries and redundancy levels.
The improvement mainly comes from \sys's design of feedback-based and incremental task dispatching algorithm.

To further understand the overall distribution of the query delay, we show the probability density function (PDF) of two queries (i.e., Q1 and Q2) in Figure~\ref{fig:eval-e2e-pdf}.
As aforementioned, we select two redundancy level--10\% and 20\%, and compare our scheduling algorithm with \texttt{OnceDispatch}.
The key observation is that our task scheduling algorithm can effectively mitigate the long-tail effects of query delay.
Indeed, when we set the redundancy to 20\%, our scheduling algorithm can achieve 2.45$\times$ and 3.43$\times$ lower 99th-MAX query delay for Q1 and Q2, respectively.
The delay reduction becomes even higher with 10\% redundancy, the 99th-MAX query delay of Q1 using \texttt{OnceDispatch} is 47,349ms and 15.15$\times$ slower than \sys.
\input{fig-eval-e2e-pdf}

\subsection{A use case: federated learning}
We also report the performance of federated learning, as an end-to-end use case for device-centric data analytics that involves multiple rounds of queries.
We test the model testing accuracy and convergence time with 30-round training.
The results are illustrated in Figure~\ref{fig:eval-fl-converge}.
With 10\% resource redundancy, \sys is able to accelerate the convergence by 1.35$\times$/1.42$\times$ as compared to \texttt{OnceDispatch} and \texttt{IncreDispatch}, respectively.
The improvement is less profound with 20\% resource redundancy, i.e., 1.10$\times$ and 1.03$\times$, respectively.
Such results demonstrate the effectiveness of \sys's task scheduling algorithm in the end-to-end, multi-round data queries.

%% file: fig-eval-e2e-query-latency.tex
\begin{figure}[t]
	\centering
	\begin{minipage}[b]{0.22\textwidth}
		\includegraphics[width=1.0\textwidth]{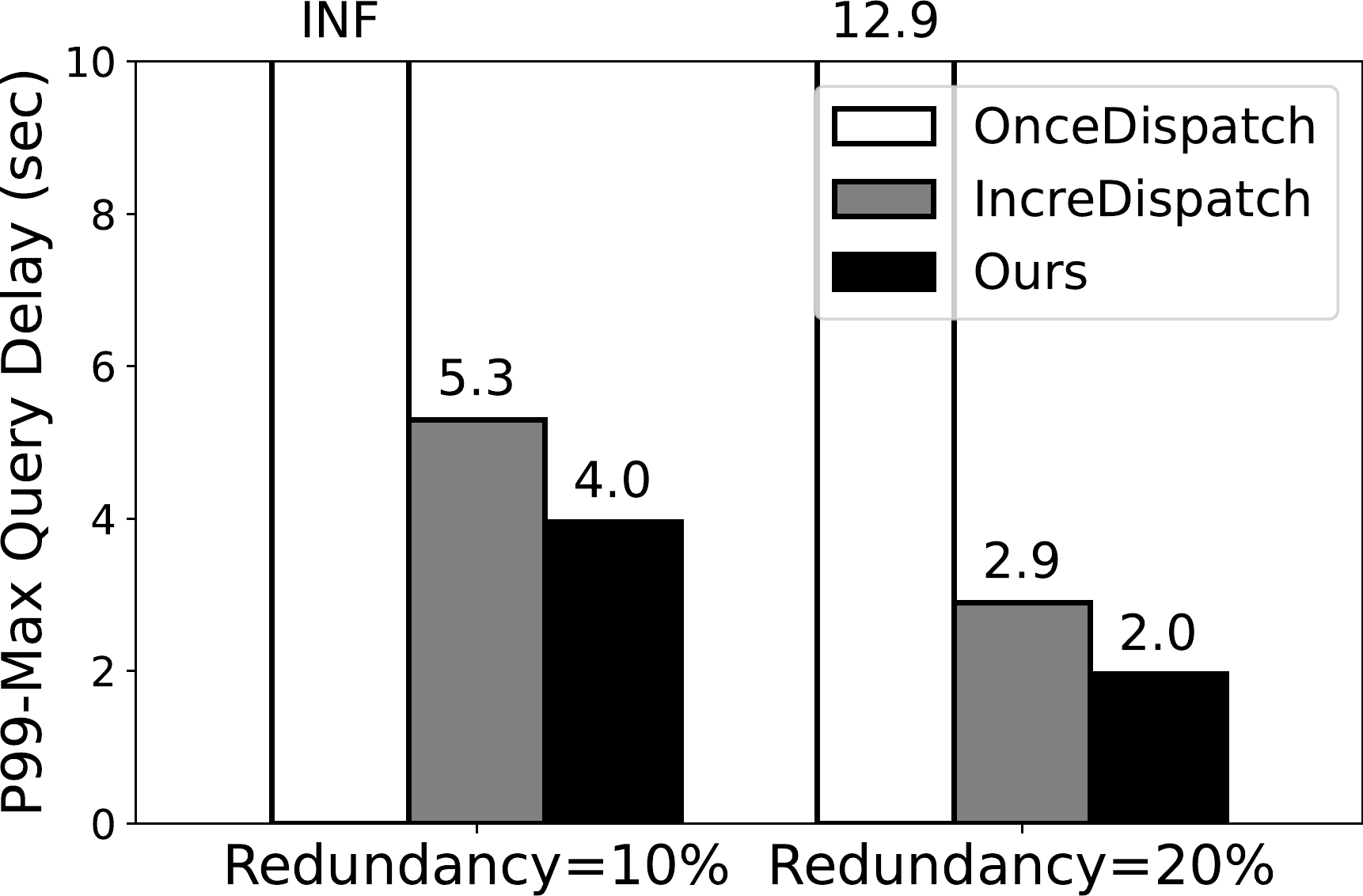}
		\subcaption{Q1}
	\end{minipage}
	~
	\begin{minipage}[b]{0.22\textwidth}
		\includegraphics[width=1.0\textwidth]{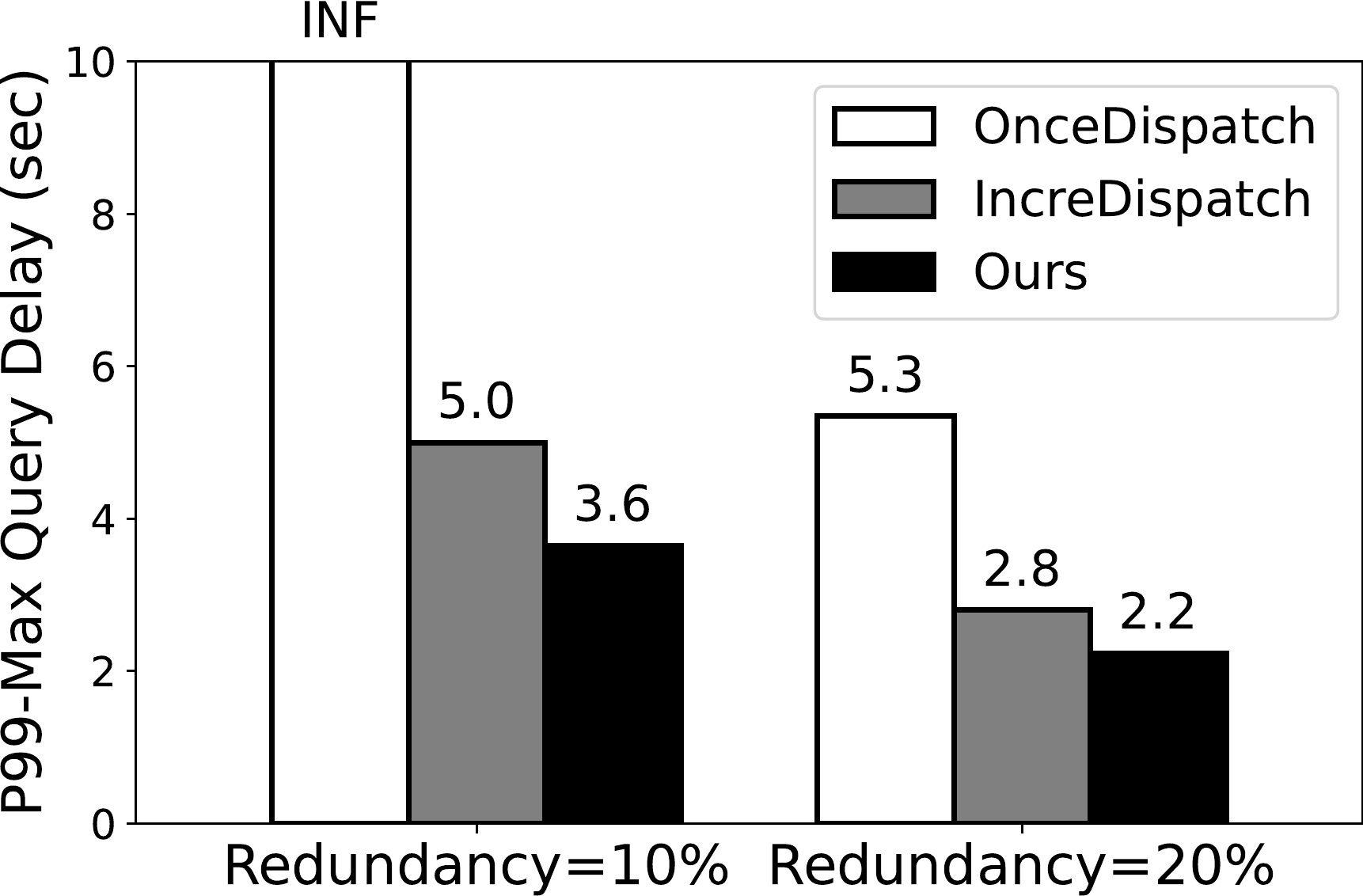}
		\subcaption{Q2}
	\end{minipage}
	
	\begin{minipage}[b]{0.22\textwidth}
		\includegraphics[width=1.0\textwidth]{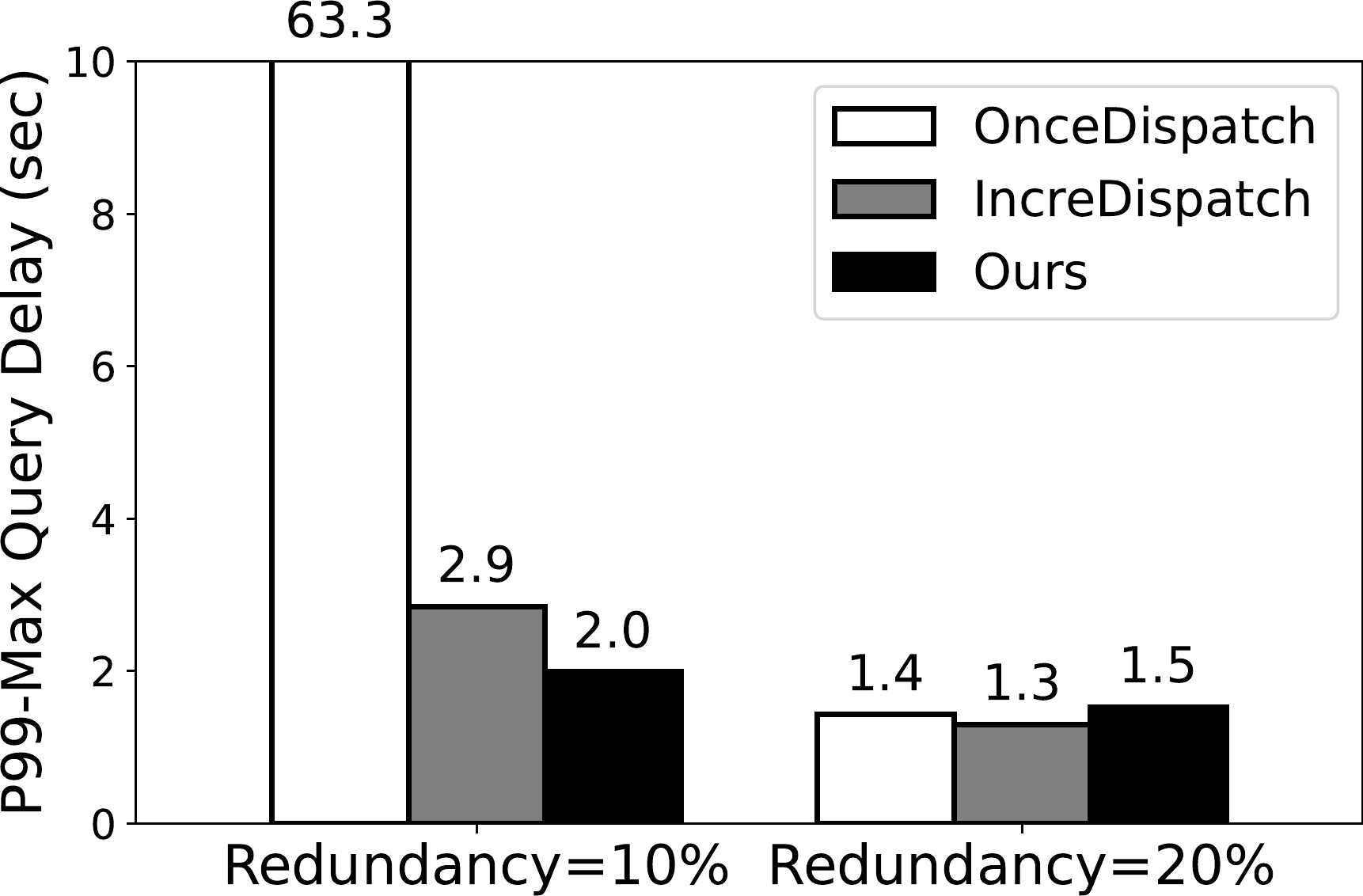}
		\subcaption{Q3}
	\end{minipage}
	~
	\begin{minipage}[b]{0.22\textwidth}
		\includegraphics[width=1.0\textwidth]{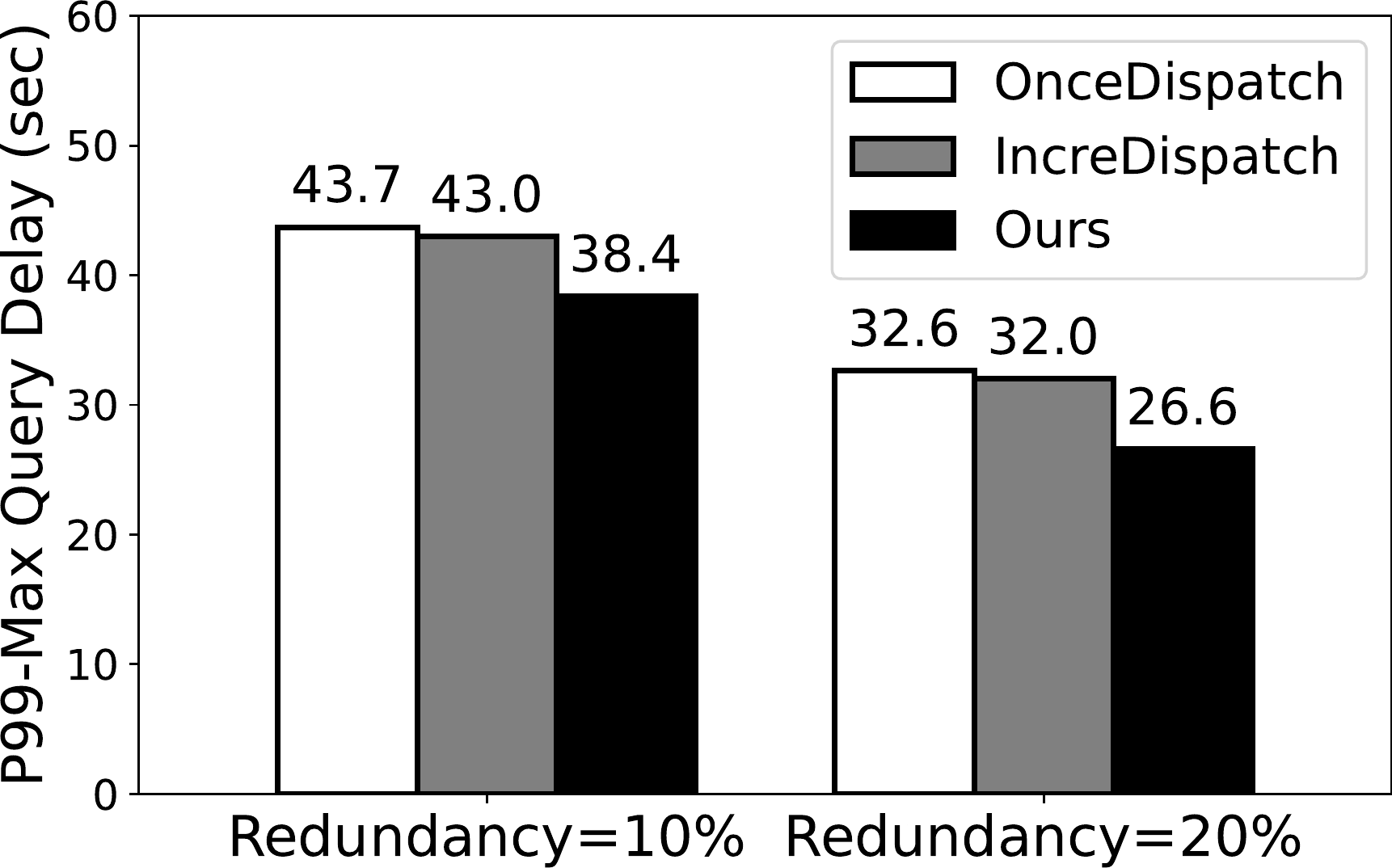}
		\subcaption{Q4}
	\end{minipage}
	\caption{The end-to-end 99th-MAX query delay under various resource redundancy.}
	\label{fig:eval-e2e-query-latency}
\end{figure}

%% file: tab-e2e-breakdown.tex
\begin{table}[t]
	\renewcommand\arraystretch{1.2}
	\centering
	\small
	\begin{tabular}{|l|l|l|l|l|}
		\hline
		\multicolumn{2}{|l|}{\multirow{2}{*}{\textbf{Query type}}} & \multicolumn{3}{l|}{\textbf{End-to-end query delay breakdown (ms)}} \\ \cline{3-5} 
		\multicolumn{2}{|l|}{} & \textbf{\begin{tabular}[c]{@{}l@{}}User-Coord\\ network delay\end{tabular}} & \textbf{\begin{tabular}[c]{@{}l@{}}Coord-side\\ pre-processing\end{tabular}} & \textbf{\begin{tabular}[c]{@{}l@{}}Task scheduling\\ (target = 100)\end{tabular}} \\ \hline
		\multirow{2}{*}{Q1} & Cold & 104 (0.8\%) & 405 (3.0\%) & 12,964 (96.2\%) \\ \cline{2-5} 
		& Warm & 29 (0.2\%) & 83 (0.6\%) & 12,890 (99.2\%) \\ \hline
		\multirow{2}{*}{Q2} & Cold & 73 (1.2\%) & 476 (8.0\%) & 5,407 (90.8\%) \\ \cline{2-5} 
		& Warm & 30 (0.6\%) & 90 (1.6\%) & 5,346 (97.8\%) \\ \hline
	\end{tabular}
	\caption{A breakdown of end-to-end query delay. ``User'': data users; ``Coord'': \gateway.}
	\label{tab:e2e-breakdown}
\end{table}

%% file: fig-eval-e2e-pdf.tex
\begin{figure}[t]
	\centering
	\begin{minipage}[b]{0.223\textwidth}
		\includegraphics[width=1.0\textwidth]{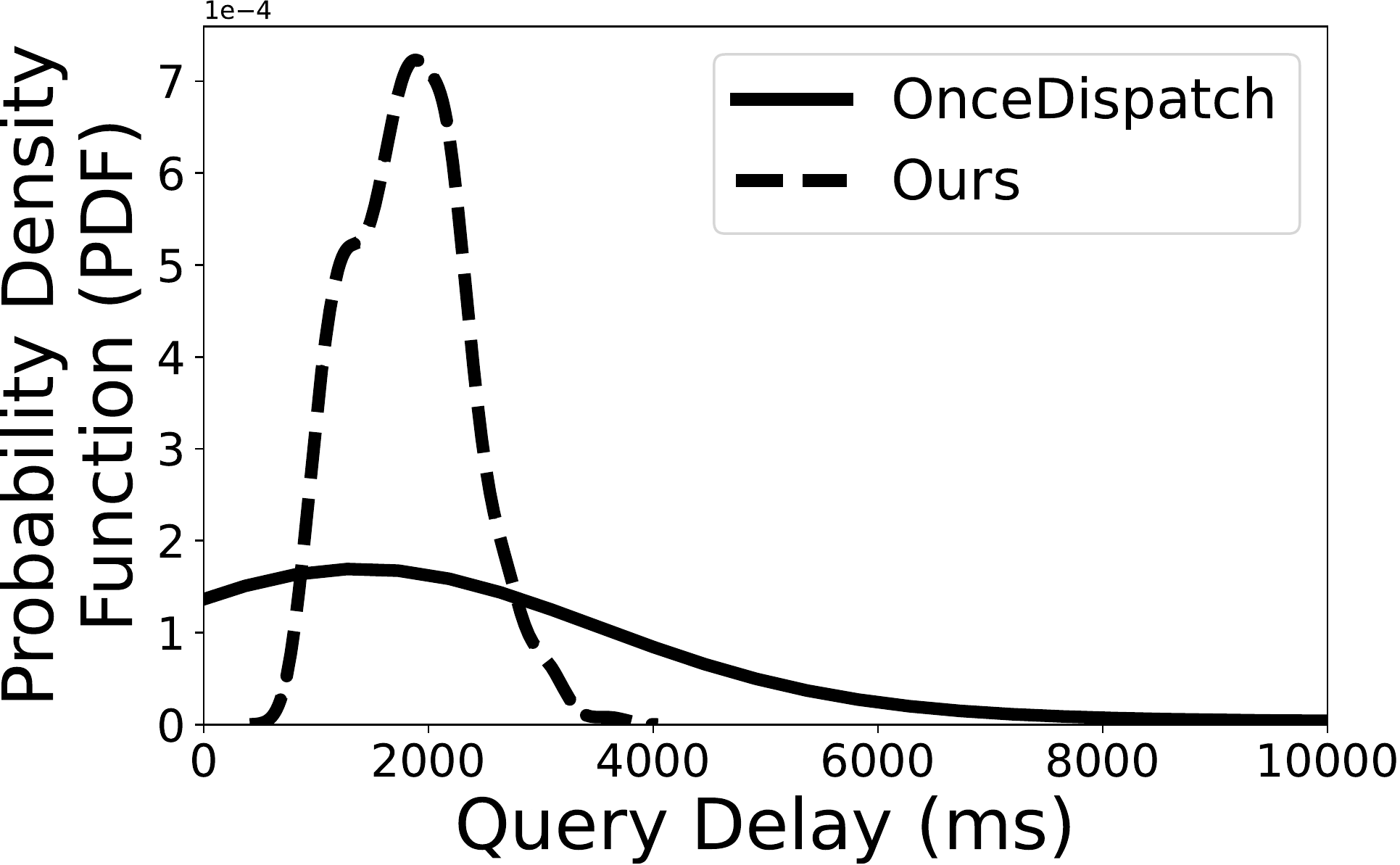}
		\subcaption{Q1, 10\% Redundancy.}
	\end{minipage}
	~
	\begin{minipage}[b]{0.223\textwidth}
		\includegraphics[width=1.0\textwidth]{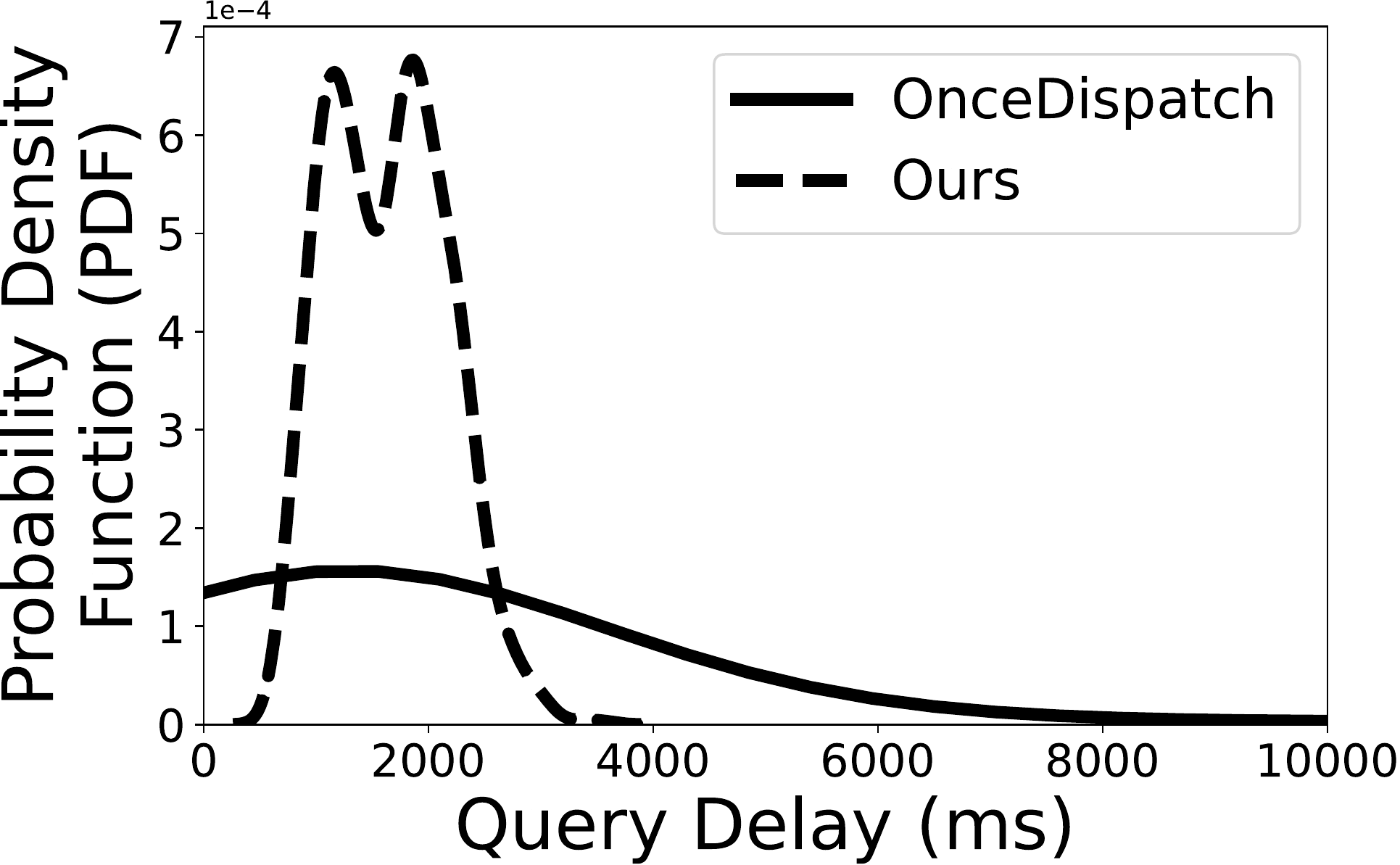}
		\subcaption{Q2, 10\% Redundancy.}
	\end{minipage}

	\begin{minipage}[b]{0.223\textwidth}
		\includegraphics[width=1.0\textwidth]{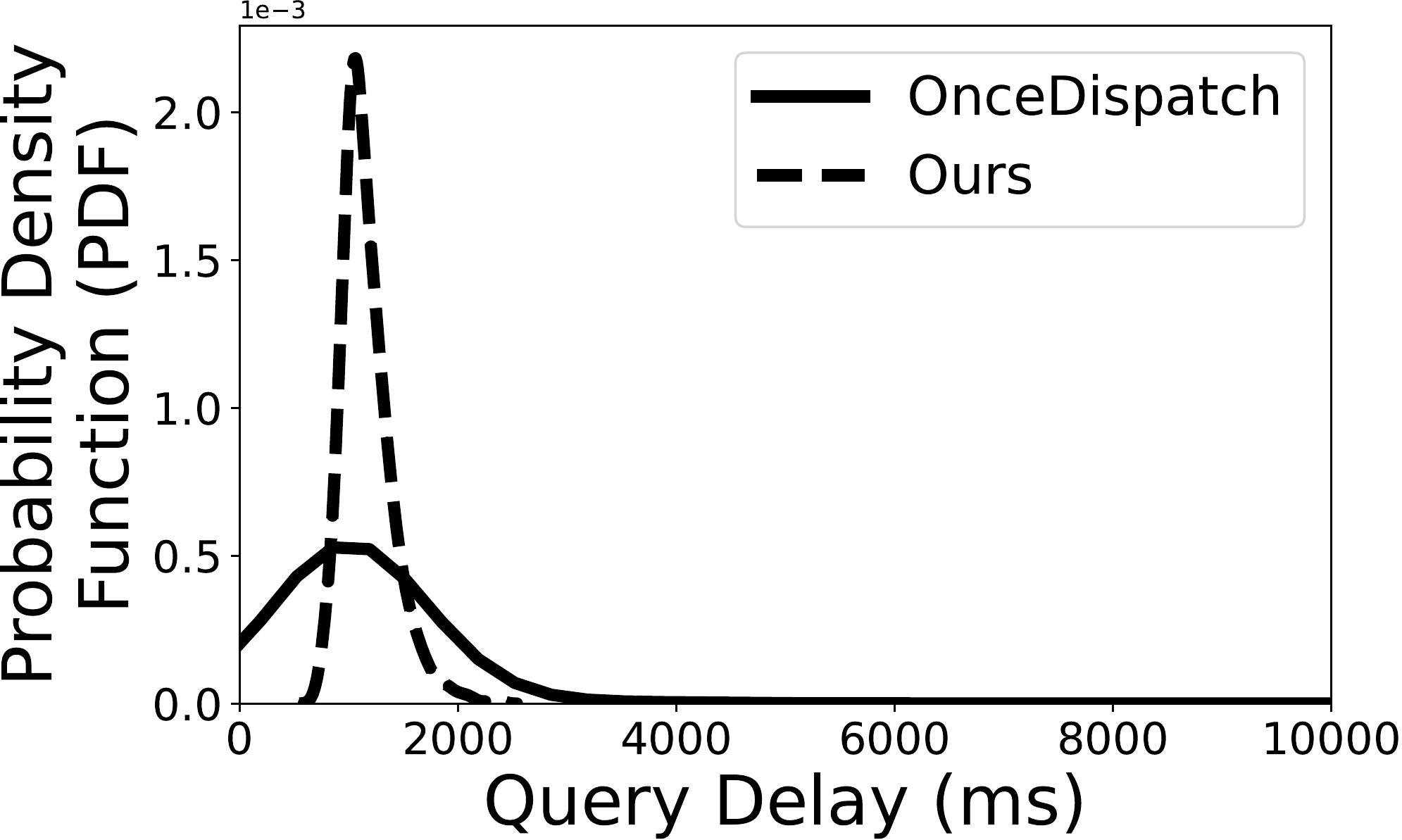}
		\subcaption{Q1, 20\% Redundancy.}
	\end{minipage}
	~
	\begin{minipage}[b]{0.223\textwidth}
		\includegraphics[width=1.0\textwidth]{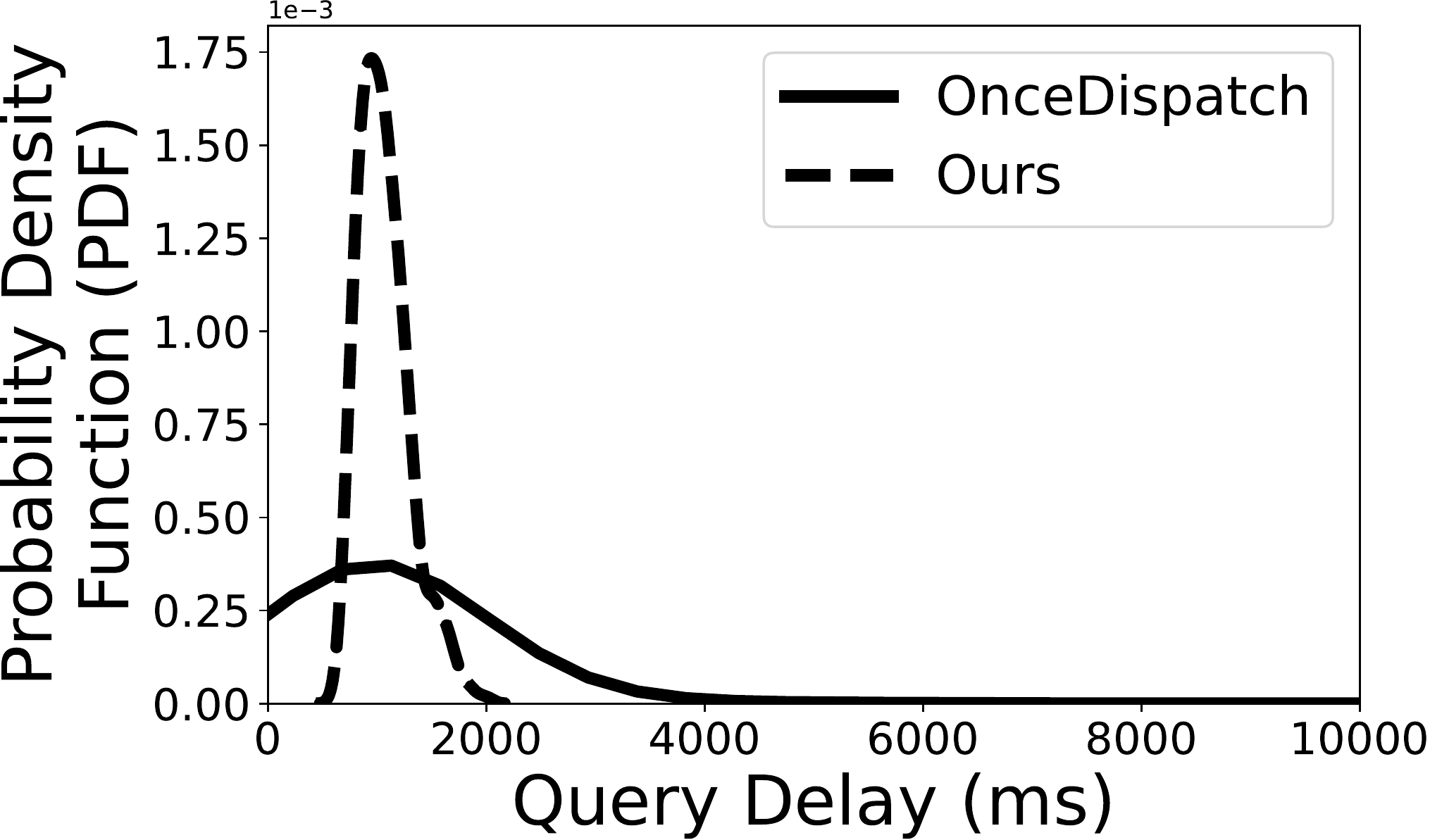}
		\subcaption{Q2, 20\% Redundancy.}
	\end{minipage}

	\caption{Probability density function (PDF) of query delay.}
	\label{fig:eval-e2e-pdf}
\end{figure}

%% file: sec-eval-overhead.tex
\subsection{System overhead}

\textbf{Standalone overhead}
We first measure the standalone runtime overhead when there is no query on the 20 \sys-enabled apps.
The experiments are performed on a rooted Meizu 16T with Qualcomm Snapdragon 855 offline.
For each app, we test the overhead for both cases when the app is running in foreground and background.
For foreground testing, we turn on the device screen and stay on the app's main page while for background we turn off the screen and wait for a minute before testing.
Each testing lasts for 5 minutes and we report the total network traffic, the average CPU usage, and the average energy consumption.
The network traffic is measured through \texttt{dumpsys netstats detail} command by filtering application \texttt{uid}.
CPU usage is read through \texttt{top} command and energy consumption is read from \texttt{sysfs} API\footnote{We keep the device fully charged during experiments and read USB power supply from \texttt{/sys/class/power\_supply/usb/}.} of Android platform.
As summarized in Table \ref{tab:apps}, \sys incurs negligible standalone overhead: at most 1.5MB APK size and 1.09/0.64 watts of energy consumption in background/foreground.
The CPU usage and network traffic increase an average of 0.06\%/0.21\% and 3.93KB/4.45KB in background/foreground, respectively. 
The overhead is primarily caused by the heartbeat messages from devices to \gateway.

\input{fig-eval-fl-converge}

\textbf{Query-time overhead}
We test the query-time overhead of \sys with the \sys-enabled input method, using the SQL query Q1 and the federated learning query Q4.
We measured device's energy consumption every 100ms and the start/end time of two queries.
We use file size to be transferred through network of each query to represent network traffic.
Results are illustrated in Figure \ref{fig:query-time-overhead}.
With the app in foreground, the real-time energy consumption can be increased by at most 5.05$\times$ for FL query and 1.68$\times$ for SQL query, as compared to standalone energy consumption.
When the app is in background, the query-time overhead can be at most 8.09$\times$ and 2.61$\times$.
In addition, \sys incurs only a few KBs network traffic in SQL queries.
An exception is that the downlink traffic of cold query in Q2 is around 1MB because a third-party library needs to be dispatched.
The network usage of the federated learning query is as high as a few MBs to exchange the DNN model.
Such nontrivial overhead motivates a task scheduling algorithm to minimize the devices being queried.

\input{fig-query-overhead}

\textbf{User survey}
Additionally, we conducted a survey on the participants involved in our field study.
We asked whether they could feel the battery consumption increased during the 2-week field study.
Among the 426 responses received, only 52 (14.6\%) reported they could perceive the increased battery consumption.
Note that during the experiments, we issued queries aggressively for data collection and system evaluation.
In reality, a device is unlikely to receive too many queries in a short period of time.

%% file: fig-eval-fl-converge.tex
\begin{figure}[t]
	\centering
	\begin{minipage}[b]{0.228\textwidth}
		\includegraphics[width=1.0\textwidth]{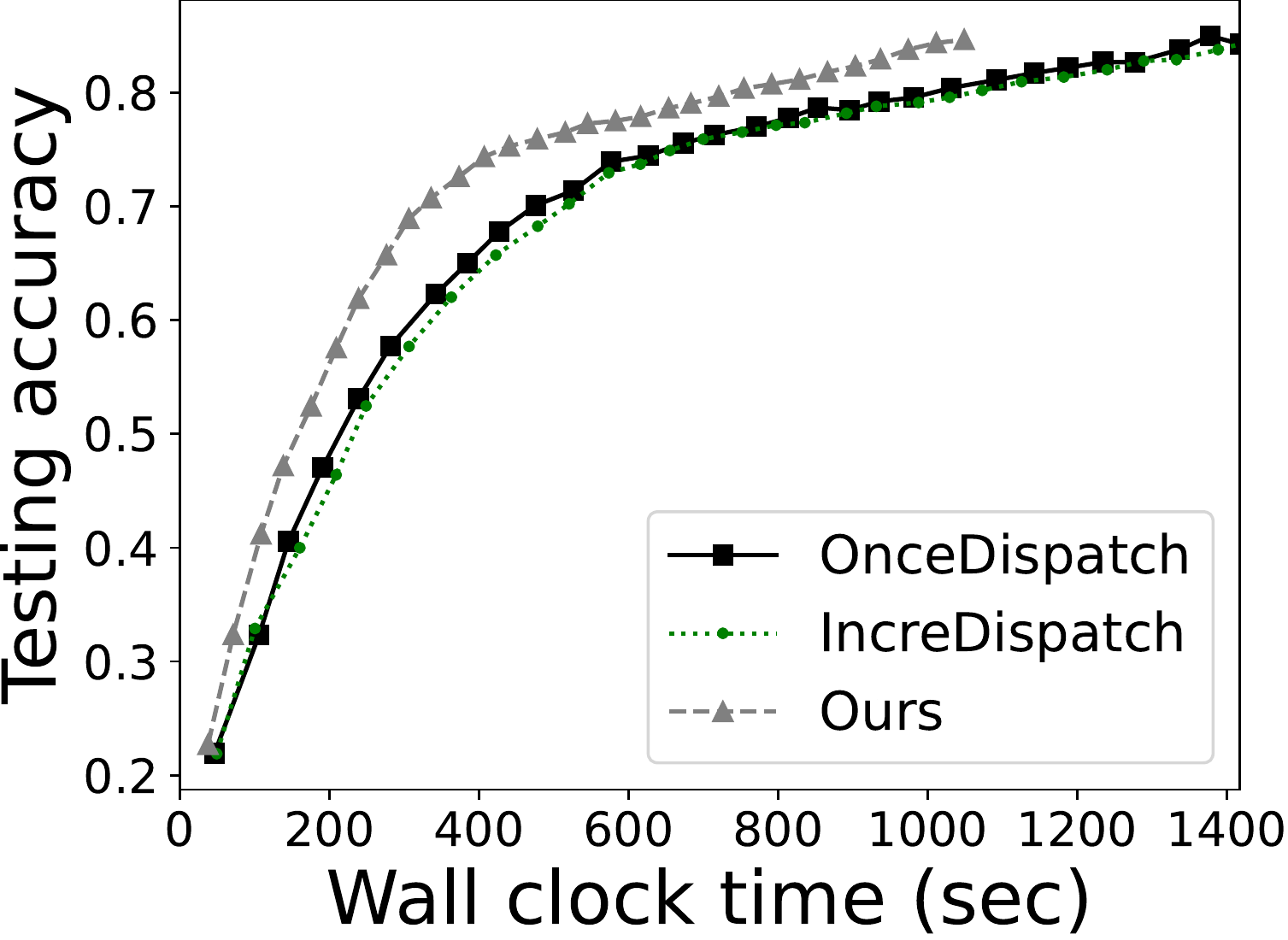}
		\subcaption{10\% redundancy}
	\end{minipage}
	~
	\begin{minipage}[b]{0.22\textwidth}
		\includegraphics[width=1.0\textwidth]{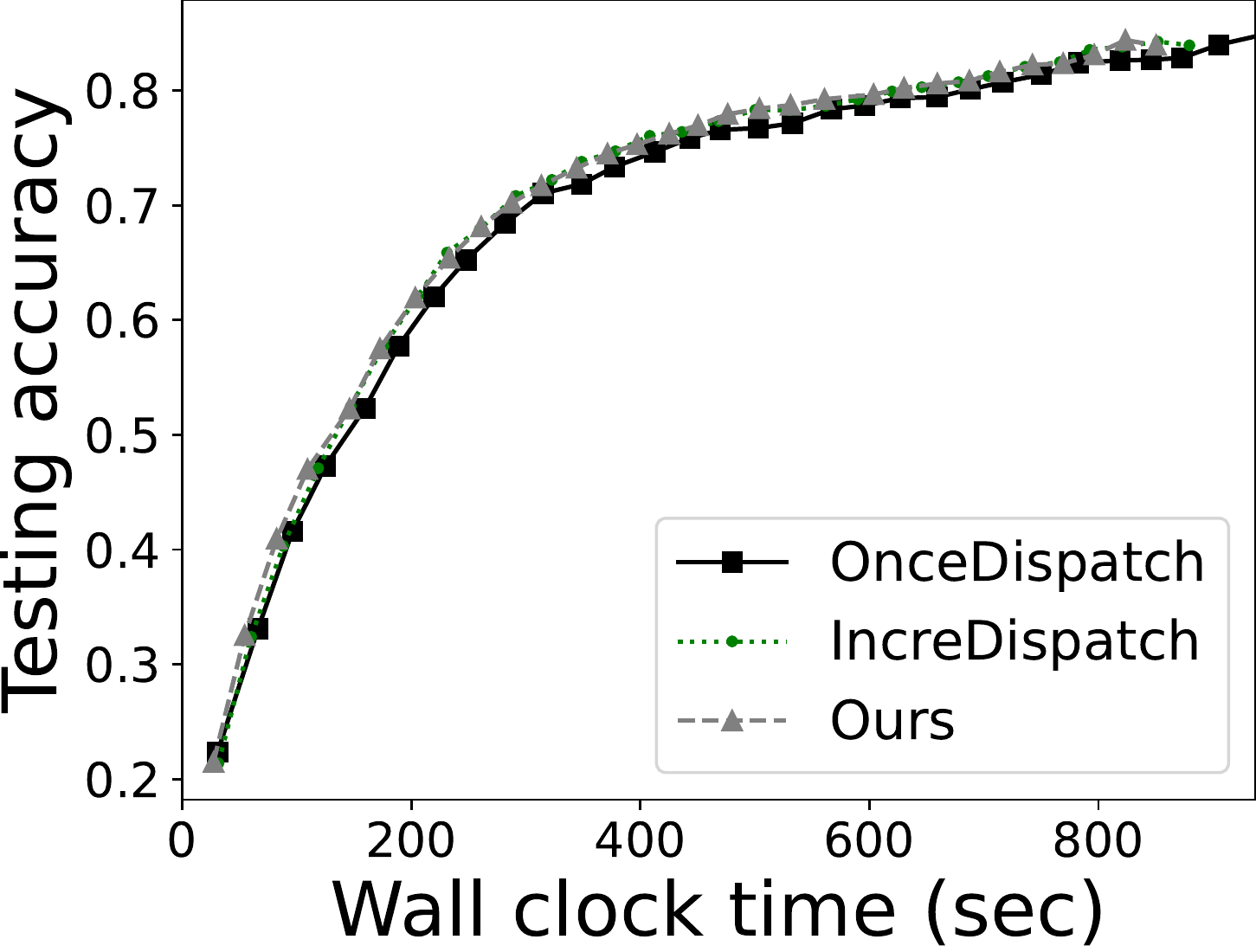}
		\subcaption{20\% redundancy}
	\end{minipage}
	\caption{The federated learning performance.}
	\label{fig:eval-fl-converge}
\end{figure}

%% file: fig-query-overhead.tex
\begin{figure}[t]
	\centering					
	\begin{minipage}[b]{0.22\textwidth}
		\includegraphics[width=1\textwidth]{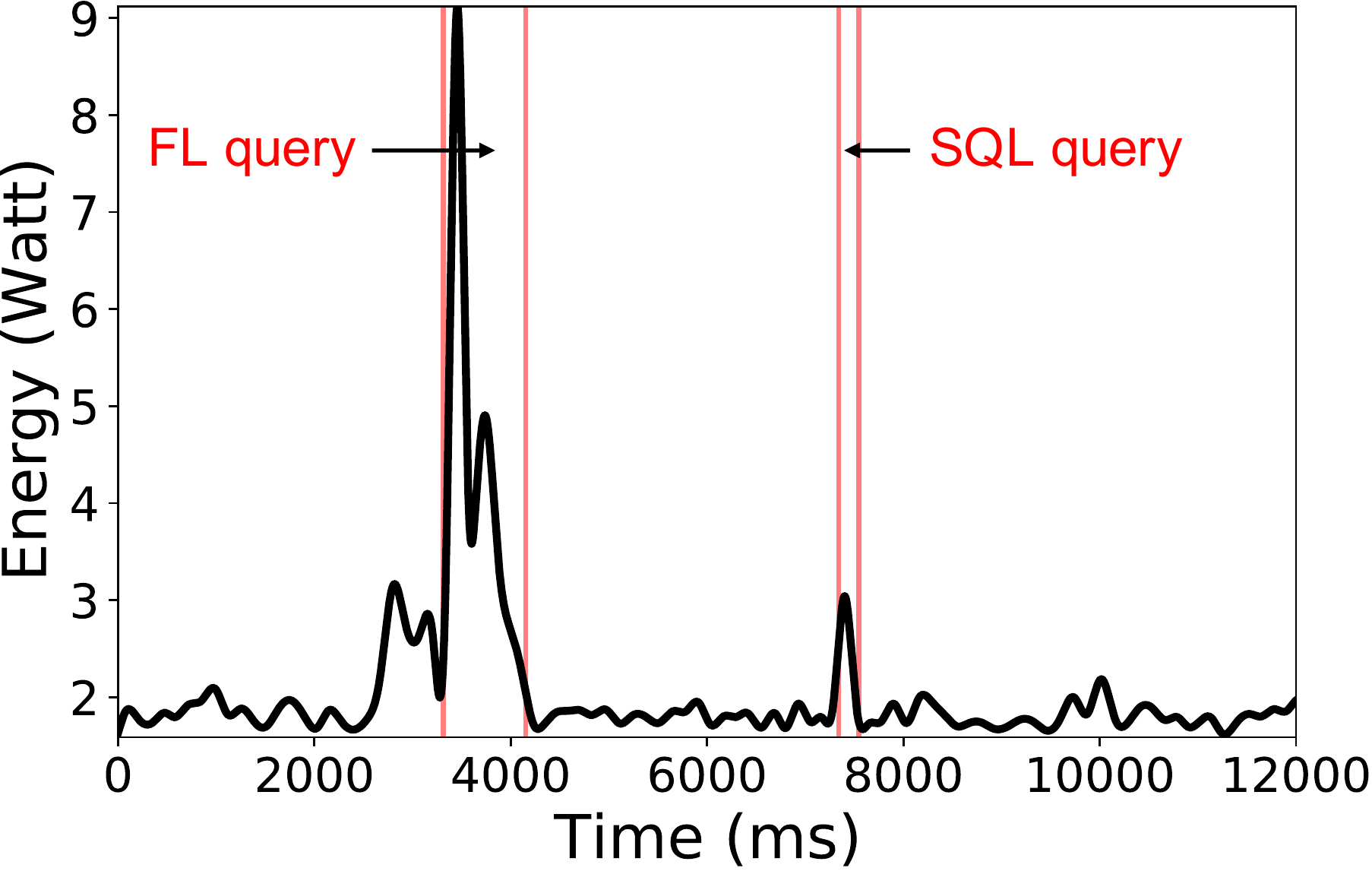}
	\subcaption{Energy at foreground}
	\end{minipage}
	\begin{minipage}[b]{0.22\textwidth}
		\includegraphics[width=1\textwidth]{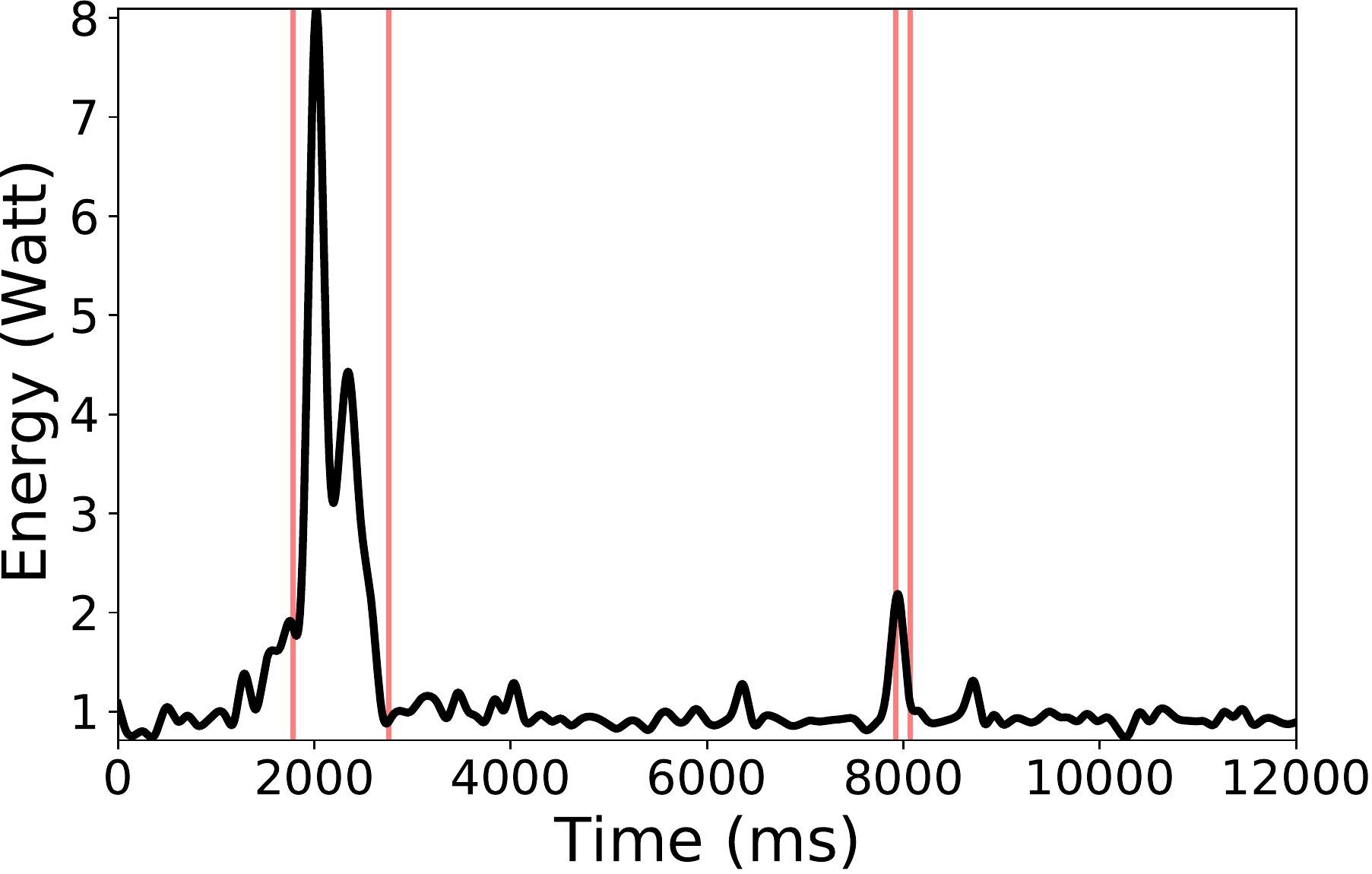}
	\subcaption{Energy at background}
	\end{minipage}
	\vspace{5pt}

	\begin{minipage}[b]{0.42\textwidth}
		\includegraphics[width=1\textwidth]{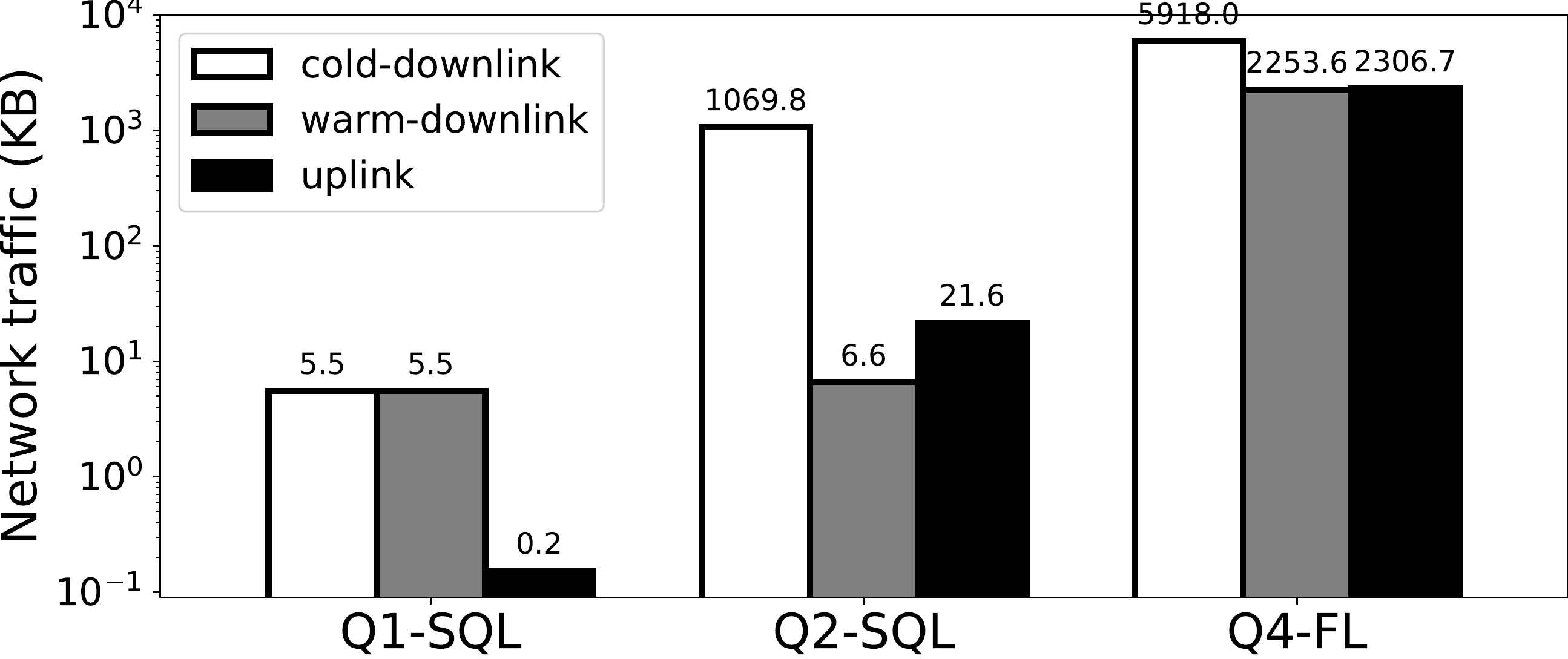}
		\subcaption{Network traffic}
	\end{minipage}
	\caption{Query-time overhead of \sys.}
	\label{fig:query-time-overhead}
\end{figure}

%% file: sec-related.tex
\input{tab-tinker-overhead}
\section{Related Work}\label{sec:related}

\noindent \textbf{Federated analytics}
As an emerging privacy-preserving data analytics paradigm, federated analytics has drawn attention from both industry~\cite{apple-dp,gboard-fl,google-playing-now,tff} and academia~\cite{infocom22-fedfpm,wang2021fedacs,hu2021feva}.
Unlike \sys that is designed for building generalized and arbitrary FA logic, previous work aim to 
(1) improve the user experience for specific mobile applications (e.g, predicting the next typed words in Gboard~\cite{gboard-fl} or identifying the music playing around the user on Pixel phones~\cite{google-playing-now})
(2) solve dedicated analytic problems, such as how to deal with data skewness in FA environments~\cite{wang2021fedacs}, or how to answer frequent pattern mining problems~\cite{infocom22-fedfpm}.
However, the primary target of \sys is to provide a uniform framework to simplify the process of designing, coding, deploying, and collecting results in FA analytics for mobile devices.
\sys also uses lightweight but efficient mechanisms described in $\S$\ref{sec:privacy} to protect data privacy in our FA scenario.
Additionally, the in-the-wild deployment shows the superior performance of \sys while others are mainly evaluated through controlled experiments.

\noindent \textbf{Data-query permission checking}
Many efforts have been made to automatically enforce privacy policies on data queries, including approaches like tracking or restricting information flows in programs~\cite{broberg2010paralocks,ferraiuolo2017verification,myers2000protecting} and proposing new programming languages~\cite{giffin2012hails,yang2012Alanguagefor,sabelfeld2003language}.
They are either tailored for specific query types (thus sacrifices programming flexibility)~\cite{sen2014bootstrapping} or require instrumentation into the system runtime (undoable for Android OS)~\cite{taintstream}.
Instead, \sys employs a hybrid method ($\S$\ref{sec:privacy}) to guarantee data privacy in best efforts.
PrivacyStreams~\cite{li2017privacystreams} protects users' privacy by providing a list of APIs for app developers to access personal data, yet only targeting single device.
\sys can leverage this work to diversify the on-device execution logic.
In the future, \sys will retrofit those techniques to further enhance privacy preservation.

\noindent \textbf{Distributed storage systems}
Partitioning data across multiple physical servers (nodes) has been a trend in recent system research due to the explosion of data volume~\cite{shipcompute_nsdi21,ford2010availability,hance2020storage}.
The recent research work on revealing the burst edge sites also facilitates this paradigm~\cite{xu2021cloud}.
For example, Ownership~\cite{wang2021ownership} automatically handles data movement for fine-grained task scheduling.
DESEARCH~\cite{desearch_osdi21} is a decentralized search engine that guarantees the integrity and privacy of search results for decentralized services.
However, similar to WAN analytics systems~\cite{pu2015low,viswanathan2016clarinet,vulimiri2015global,vulimiri2015wanalytics,sol-nsdi,rabkin2014aggregation,zhang2018awstream}, they focus on data placement and movement across servers instead of mobile devices, therefore ignoring the constrained hardware resource and the potentially malicious data queries.

\noindent \textbf{Hot-fix libraries}
Android hot-fix libraries are mainly used to fix application bugs immediately without reinstallation or rebooting~\cite{noauthor_tencenttinker_2021,noauthor_andfix_2021,noauthor_robust_2021,noauthor_alibabadexposed_2021} such as Tinker~\cite{noauthor_tencenttinker_2021}.
Compared with them, \sys provides unified and flexible programming interfaces directly to data users instead of app developers, and \sys is much faster in task compiling as it decouples the analytics logic from app logic as experimentally shown in $\S$\ref{sec:eval-usbability}.

\noindent \textbf{Crowdsourcing platforms}~\cite{hirth2011anatomy,van2012designing,yan2009mcrowd} enable developers or researchers to obtain data from a relatively large, geo-distributed group of participants.
For example, Funf~\cite{funf} is an open sensing framework that provides reusable functionalities to app developers for data collection, uploading, and configuration.
Those frameworks are mostly built for data collection instead of querying systems, thus do not address the performance and privacy challenges as \sys does.

%% file: tab-tinker-overhead.tex
\begin{table}[t]
    \renewcommand\arraystretch{1.2}
    \centering
	\small
    \begin{tabular}{|c|c|c|c|c|}
        \hline
        \multicolumn{1}{|l|}{}           & \textbf{\begin{tabular}[c]{@{}c@{}}Deck\\ (SQL)\end{tabular}} & \textbf{\begin{tabular}[c]{@{}c@{}}Tinker\\ (SQL)\end{tabular}} & \textbf{\begin{tabular}[c]{@{}c@{}}Deck\\ (Image)\end{tabular}} & \textbf{\begin{tabular}[c]{@{}c@{}}Tinker\\ (Image)\end{tabular}} \\ \hline
        \textbf{Dispatch Size (KB)}      & 2.53                                                          & 5.04                                                            & 407                                                             & 151                                                               \\ \hline
        \textbf{Compile Time (s)}        & 1.24                                                          & 53.72                                                           & 1.13                                                            & 156.43                                                            \\ \hline
        \textbf{Network Time (s)}        & 0.029                                                         & 0.031                                                           & 0.104                                                           & 0.065                                                             \\ \hline
        \textbf{Exec Time (s)}           & 0.125                                                         & 1.951                                                           & 0.272                                                           & 2.219                                                             \\ \hline
        \textbf{Total (s)}               & 1.394                                                         & 55.702                                                          & 1.506                                                           & 158.714                                                           \\ \hline
        \end{tabular}
    \caption{Comparison between \sys and Tinker with SQL and image processing queries.}
    \label{tab:tinker-overhead}
\end{table}

%% file: sec-discuss.tex
\section{Discussion}

\textbf{Privacy guarantee}
Powered by our self-designed APIs listed in Table~\ref{tab:APIs}, \sys currently aims to allow data users to reveal the statistical patterns of the distributed databases on clients.
At the coordinator side, the privacy design forces data users to perform mandatory aggregation operation, which helps reduce the privacy leakage but also narrows the supported range of query types.
To balance the privacy and programming flexibility, we can loose the mandatory aggregation constraint at the coordinator side, but enable local differential privacy~\cite{erlingsson2014rappor} on each devices.
Such a design also makes data users hard to get the concrete individual device data, but possible to operate on pre-aggregation data with a little data utility cost incurred by local differential privacy.

\sys can be potentially integrated with other advanced privacy and security techniques.
For example, we can leverage the hardware-based solutions (e.g., ARM TrustZone and Intel SGX) to avoid data leakage from both device-side and server-side~\cite{mo2021ppfl}.
Isolating function execution environments is commonly used in cloud to power serverless computing by re-purposing the existing container platforms.
However, container-based virtualization techniques~\cite{andrus2011cells,song2021towards} fail in their coarse-grained isolating level (e.g., the overall Android filesystem), and thus are unable to apply to \sys.
Recent work on isolating individual functions~\cite{wanninger2022isolating} may be a promising solution.

\textbf{Usability enhancement}
\sys asks data users to specify the number of devices to be queried.
In reality, it could be challenging for data users to give a ``proper'' device number that exhibits satisfactory accuracy and also low resource expenditure.
We plan to extend \sys to accept the parameter as a confidence interval (e.g., 10\%) and confidence level (e.g., 95\%).
For example, a data user may want to know the range (X, X+y) of the number of a song being played within a week.
\sys will keep querying devices till the specified confidence is achieved.
Confidence interval is extensively used in online query refinement~\cite{xu2020approximate,agarwal2013blinkdb} and can further enhance the usability of \sys.
Enabling such query mode requires \sys to revise its central \gateway and the device scheduling algorithm.

%% file: sec-conclusions.tex
\section{Conclusions}\label{sec:conclusions}

In this work, we propose a unified framework \sys for elastic and efficient on-device federated analytics.
\sys facilitates data users to focus on the analytics logic while handles the device-specific tasks underneath.
\sys also incorporates novel techniques to guard data privacy and trades off query delay with analytics resource expenditure on devices.
The effectiveness of \sys is demonstrated through both large-scale field deployment and offline microbenchmarks.